\documentclass[12pt]{article}
\usepackage{amsmath, amssymb, amsfonts, amsthm}
\usepackage{mathrsfs}
\usepackage{natbib,graphicx,setspace,lscape,longtable}
\usepackage{mathrsfs,amsmath,amsthm,amssymb,color}
\usepackage{natbib,epsfig,graphicx,pdfpages}
\usepackage{rotating}
\usepackage{subfigure}
\usepackage{enumerate}
\usepackage{graphicx}
\usepackage{algorithm}
\usepackage{algpseudocode}
\usepackage{url}
\usepackage{hyperref}
\usepackage[table]{xcolor}
\usepackage{colortbl}
\usepackage{adjustbox}
\bibpunct{(}{)}{;}{a}{,}{,}
\usepackage[compact]{titlesec}
\titlespacing{\section}{0pt}{*0}{*0}
\titlespacing{\subsection}{0pt}{*0}{*0}

\usepackage{longtable}
\usepackage{multirow}
\usepackage{float}
\usepackage{lscape}
\usepackage{algorithmicx}
\usepackage{authblk}
\usepackage{makecell}

\usepackage{booktabs}

\hypersetup
{   
colorlinks=true,
 linkcolor=blue,
urlcolor={red},
citecolor=blue
}

\usepackage{enumitem}
\setlist{itemsep=.01em}
\setlist{topsep=.5em}
\allowdisplaybreaks
\newtheorem{example}{Example}
\newtheorem{theorem}{Theorem}
\newtheorem{lemma}{Lemma}
\newtheorem{proposition}{Proposition}
\newtheorem{remark}{Remark}

\newtheorem{assumption}{{Assumption}}

\def\beq{\begin{equation}}
\def\eeq{\end{equation}}
\def\beqr{\begin{eqnarray}}
\def\eeqr{\end{eqnarray}}
\def\beqrs{\begin{eqnarray*}}
\def\eeqrs{\end{eqnarray*}}
\def\bet{\begin{theorem}}
\def\eet{\end{theorem}}
\def\bel{\begin{lemma}}
\def\eel{\end{lemma}}
\def\bep{\begin{proposition}}
\def\eep{\end{proposition}}
\def\bg{\begin{figure}[tbph]\begin{center}}
\def\eg{\end{center}\end{figure}}

\def\bc{\begin{center}}
\def\ec{\end{center}}

\def\wt{\widetilde}

\newcommand{\ve}{{\varepsilon}}
\renewcommand{\epsilon}{{\ve}}
\renewcommand{\hat}{\widehat}

\def\wt{\widetilde}
\renewcommand{\tilde}{\wt}

\numberwithin{equation}{section}

\begin{document}



\def\spacingset#1{\renewcommand{\baselinestretch}%
{#1}\small\normalsize} \spacingset{1}


{
 
\title{Forward Regression via Gram–Schmidt Orthogonalization for Ultra-High Dimensional Linear Models}
	\author{
		
Jialuo Chen$^1$, Zhaoxing Gao$^2$\thanks{Corresponding author. Gao thanks the National Natural Science Foundation of China (NSFC) for financial support under the grant number 72573029.}, Yifan Jiang$^3$, and Ruey S. Tsay$^4$\\
$^1$School of Management, Zhejiang University\\
$^2$School of Mathematical Sciences and School of Economics and\\ Management,
University of Electronic Science and Technology of China\\
$^3$Perelman School of Medicine, University of Pennsylvania\\
$^4$Booth School of Business, University of Chicago
}	
 \date{}

\maketitle
}

\begin{abstract}{\normalsize
Forward regression is a classical and effective tool for variable screening in ultra-high dimensional linear models, but its standard projection-based implementation can be computationally costly and numerically unstable when predictors are strongly collinear. Motivated by this limitation, we propose an orthogonalized forward regression procedure, implemented recursively through Gram–Schmidt updates, that ranks predictors according to their unique contributions after removing the effects of variables already selected. This approach preserves the interpretability of forward regression while substantially reducing the cost of repeated projections. We further develop a path-based model size selection rule using statistics computed directly from the forward sequence, thereby avoiding cross-validation and extensive tuning. The resulting method is particularly well suited to settings in which the number of predictors far exceeds the sample size and strong collinearity renders the conventional forward fitting ineffective. Theoretically, we derive the optimal convergence rate for the proposed Gram–Schmidt forward regression, thereby extending existing results for projection-based forward regression, and further show that it enjoys sure screening property and variable selection consistency under suitable conditions. Simulation studies and empirical examples demonstrate that it provides a favorable balance among computational efficiency, numerical stability, screening accuracy, and predictive performance, especially in highly correlated ultra-high dimensional settings.
}  
\end{abstract}

\noindent%
				
\vfill

\newpage
\spacingset{1.75} 


\abovedisplayskip=0.1pt
\belowdisplayskip=0.1pt


		\section{Introduction}
High-dimensional regression models have become increasingly important in modern statistics, econometrics and empirical applications. 
They facilitate the analysis of the relationship between a response variable and multiple predictors {via} 
\begin{equation}
    y_{t} = \alpha + \sum_{j=1}^{p} \beta_{j} x_{t j} + \varepsilon_{t}, \quad t = 1, \ldots, n,
    \label{model}
\end{equation}  
where \(y_{t}\) is the response variable, \(x_{t 1}, \ldots, x_{t p}\) are the feature variables, \(n\) is the sample size, and \(\varepsilon_{t}\) is a mean-zero disturbance term. Modern regression analysis increasingly encounters ultra-high dimensional 
datasets in which the number of 
 predictors far exceeds the sample size. In such settings, direct regression fitting is often unstable or infeasible because the sample covariance matrix 
may be ill-conditioned and the computational burden of large-scale model exploration can be substantial (\citet{ledoit2004well}; \citet{10.1214/009053607000000758}). In many important applications, this difficulty is exacerbated by strong multicollinearity among predictors, as in macroeconomic forecasting, financial signal extraction, genomics, and other large-scale empirical problems. The joint presence of ultra-high dimensionality and high multicollinearity therefore creates a demanding screening problem, calling for new methods that are computationally scalable, numerically stable, and effective at retaining the relevant 
predictors without selecting an overly dense model.

\par 

In high-dimensional settings, where the number of predictors may grow much faster than the sample size, penalized regression methods have become a standard tool for variable selection. Popular approaches, such as Lasso, SCAD, Bridge, and Elastic Net, impose regularization to encourage sparse representations and improve interpretability (\citet{tibshirani1996regression}; \citet{fan2001variable}; \citet{huang2008asymptotic}; \citet{zou2005regularization}). Despite their success, these methods can face important difficulties when the predictor dimension is extremely large and the covariates are strongly correlated. In particular, nonconvex penalization methods may involve substantial computational complexity, while convex procedures can suffer from unstable variable selection or reduced screening accuracy under high multicollinearity and weak signals. These limitations have motivated the study of alternative approaches for ultra-high dimensional variable screening that seek to balance computational scalability, effective dimension reduction, and reliable recovery of relevant predictors. Among such alternatives, greedy forward procedures are particularly appealing because they construct models sequentially and provide a transparent and scalable mechanism for variable ranking.

\par


Greedy forward selection procedures construct models sequentially by adding predictors according to a predefined ranking rule. Representative examples include $L_2$-Boosting (\citet{Peter2006}), the Orthogonal Greedy Algorithm (OGA) (\citet{ing2020model}; \citet{ingStepwiseRegressionMethod2011}), and forward regression procedures (\citet{donoho2006breakdown}; \citet{barron2008approximation}; \citet{wang2009forward}). These methods are computationally attractive and often perform well in ultra-high dimensional settings. However, existing greedy algorithms typically rely on either marginal correlations or repeated projection steps to rank candidate predictors. Marginal-correlation-based methods, such as OGA and $L_2$-Boosting, may misidentify relevant variables when predictors are strongly correlated, because marginal correlations can be distorted by collinearity. Projection-based forward regression methods evaluate incremental contributions more directly, but 
 their repeated projection operations are computationally expensive; moreover, when candidate predictors are nearly collinear with previously selected variables, these updates can become numerically unstable.

  To address these computational and stability issues, this paper studies an efficient Gram–Schmidt implementation of forward regression for ultra-high dimensional linear models. Rather than repeatedly recomputing projections onto the augmented selected set, we update the forward path through recursive orthogonalization and rank candidate predictors according to their orthogonalized incremental contributions. This preserves the forward-regression logic of selecting variables by additional explanatory power, while making the procedure substantially more stable and computationally attractive in highly correlated designs. The procedure is run for $K_n$ steps, and the final model is then determined by a data-driven model size selection rule constructed along the orthogonalized forward path.   We establish theoretical properties of the proposed procedure in ultra-high dimensional settings where the number of predictors may grow exponentially with the sample size. 
{Specifically, under some suitable conditions, we show that the proposed Gram--Schmidt forward regression (GSFR) attains the optimal convergence rate, comparable to that of the orthogonal greedy algorithm, and enjoys the sure screening property.} In addition, we prove that the proposed model size selection rule consistently identifies the optimal model size along the forward selection path.

\par

Our approach is related to orthogonal greedy algorithms such as those studied in \citet{ingStepwiseRegressionMethod2011}, but it differs fundamentally in the way candidate variables are evaluated. Instead of ranking predictors by their marginal correlations with the current residual, we evaluate their orthogonalized incremental contributions conditional on the variables already selected. This distinction becomes particularly important under strong multicollinearity, where marginal ranking can be unstable and potentially misleading. From a computational standpoint, the proposed procedure can be viewed as a Gram--Schmidt implementation of forward regression that avoids repeated augmented projections, making it especially appealing in highly collinear ultra-high dimensional settings.

The contribution of this paper is threefold. First, we develop a Gram–Schmidt implementation of forward regression that evaluates variables through orthogonalized incremental contributions and is computationally more attractive than repeated projection-based updates in strongly collinear designs. Second, we propose a simple path-based stopping rule that can be computed directly along the forward sequence and helps identify a parsimonious model without cross-validation {or tuning}. Third, we show that these computational and practical improvements do not come at the cost of theoretical reliability: 
{The resulting procedure attains the optimal convergence rate and, under suitable conditions, is guaranteed to possess the sure screening property and variable selection consistency.}
Our simulations and empirical examples are designed to reflect the intended use of the method, namely large-scale regression problems in which both multicollinearity and computation are first-order concerns. 

The rest of the paper is organized as follows. Section \ref{sec2} briefly reviews the Orthogonal Greedy Algorithm (OGA) and the classical projection-based Forward Regression (FR), presents counterexamples to motivate our proposed GSFR approach, and provides a detailed introduction to the proposed algorithm. Section \ref{sec3} investigates the theoretical properties of GSFR. Section \ref{sec4} evaluates the numerical performance of the proposed approach through simulations and two real data examples, including comparisons with some existing methods. Section \ref{sec5} concludes. All proofs and derivations for the asymptotic results are provided in an online Appendix.

		\section{Methodology}\label{sec2}

        In this section, we first review the Orthogonal Greedy Algorithm (OGA)  {of} 
\citet{ingStepwiseRegressionMethod2011} and illustrate a setting in which marginal-correlation-based ranking may fail under strong predictor dependence. This motivates us to study a Gram–Schmidt implementation of forward regression that ranks predictors by their orthogonalized incremental contributions and avoids repeated augmented projection steps. In what follows, we use projection-based FR (PBFR) to refer to the conventional implementation studied by Wang (2009) in high-dimensional settings, in contrast to the Gram–Schmidt implementation studied here. We present an additional example illustrating that standard projection-based implementations of FR  may become numerically unstable when predictors are nearly collinear. This illustrates the appeal of the GSFR implementation, which retains the orthogonalized incremental ranking logic while avoiding repeated augmented projections through recursive Gram–Schmidt updates. Finally, we describe the implementation of GSFR and introduce a new path-based rule for model size selection.

We {start with some notation.} Let \( y_t \) denote the observed target variable, and \( x_{t1}, \ldots, x_{tp} \) denote the observed predictors,  where \( t = 1, \ldots, n \).
 For simplicity, we set the intercept \( \alpha = 0 \) in Model (\ref{model}). The observed data are centered by replacing \( y_t \) with \( y_t - \bar{y} \) and \( x_{tj} \) with \( x_{tj} - \bar{x}_j \), where \( \bar{x}_j = n^{-1} \sum_{t=1}^n x_{tj} \) and \( \bar{y} = n^{-1} \sum_{t=1}^n y_t \). The data are assumed to be square-integrable random variables with zero means, satisfying \( E(x_{tj}^2) = \sigma_{j}^{2} \), for $j=1,\ldots,p$. Let \( \mathbf{x}_t = (x_{t1}, \ldots, x_{tp})^\prime \) denote the data vector. We assume an independent and identically distributed (i.i.d.) setting, where \( (\varepsilon_t, {\mathbf{x}}^\prime_t  )\) are i.i.d. and \( \varepsilon_t \) is independent of \( \mathbf{x}_t \). To distinguish between the population and sample versions of the model, let \( (\varepsilon, \mathbf{x}^\prime) \) denote an independent replication of \( (\varepsilon_t, {\mathbf{x}}^{\prime}_t) \). At the population level, the model is represented as \( y(\mathbf{x}) = \boldsymbol{\beta}^\prime \mathbf{x} \) and \( y = y(\mathbf{x}) + \varepsilon \). Let \( y_J(\mathbf{x}) \) denote the best linear predictor of \( y(\mathbf{x}) \) based on \( \{x_j : j \in J\} \), where \( J \) is a subset of \( \{1, \ldots, p_n\} \). Similarly, let \( \hat{y}_{t; J} \) denote the fitted value of \( y_t \) when \( \mathbf{Y} = (y_1, \ldots, y_n)^\prime \) is projected onto the linear space spanned by \( \mathbf{X}_j \), with \( \mathbf{X}_j = (x_{1j}, \ldots, x_{nj})^\prime \), for \( j \in J \neq \emptyset \). If \( J = \emptyset \), we set \( \hat{y}_{t; J} = 0 \). Let \( J_k \) and \( \hat{J}_k \) denote the sets of variable indices selected at the end of stage \( k \) under the population and sample versions of GSFR or OGA, respectively, where \( J_k = \{j_1, \ldots, j_k\} \) corresponds to the set of selected indices in the population model, and \( \hat{J}_k = \{\hat{j}_1, \ldots, \hat{j}_k\} \) represents the selected indices in the sample model. The \(\ell_1\)-norm of a vector \(\boldsymbol{\nu} = (\nu_1, \dots, \nu_k)^\prime\) is defined as
\(
\|\boldsymbol{\nu}\|_1 = \sum_{j=1}^k |\nu_j|.
\)



\label{section2}
\subsection{Orthogonal Greedy Algorithm (OGA)}

The OGA 
utilizes an iterative selection procedure based on the marginal contributions of variables to the residuals {of $y_t$.} At each iteration, the next variable is chosen according to the following criteria:
\[
j_{k+1} = \arg \max_{1 \leq j \leq p} |\mu_{J_k, j}|, \quad
\hat{j}_{k+1} = \arg \max_{1 \leq j \leq p} |\hat{\mu} _{\hat{J}_k, j}|.
\]
Here, $\mu_{J_k, j}$ and $\hat{\mu} _{\hat{J}_k, j}$ represent the marginal contributions of variables at the population and sample levels, respectively. For \( i \notin J \), these contributions are defined as

\begin{equation}
    \mu_{J, i}=E\left[\left\{y(\mathbf{x})-y_{J}(\mathbf{x})\right\} x_{i}\right], \quad 
    \hat{\mu}_{J, i}=\frac{n^{-1} \sum_{t=1}^{n}\left(y_{t}-\hat{y}_{t ; J}\right) x_{t i}}{\left(n^{-1} \sum_{t=1}^{n} x_{t i}^{2}\right)^{1 / 2}},
    \label{ref1}
\end{equation}
where the population version of OGA corresponds is a special case of the Weak Orthogonal Greedy Algorithm (WOGA) in \citet{temlyakov2000weak}.

The values \( \mu_{J, i} \) and \( \hat{\mu}_{J, i} \) measure the marginal contribution of the variable \( x_i \) to the residuals, reflecting its potential to explain the remaining variation in the response. At each iteration, OGA selects the variable that minimizes the residual sum of squares (RSS) when the residual is regressed onto the candidate variable. 

\subsection{A Motivating Example Where OGA May Fail
}\label{sec2.2}

In this section, we construct an example with three predictors to evaluate the performance of OGA in variable selection when the predictors 
may be highly correlated. The high correlations pose a challenging problem 
for OGA to select the true predictors. 
{We examine the selection outcomes of OGA after two iterations.}


\begin{example}
{Consider} the model 
\begin{align}
    y &= \beta x_{1} + x_{2} + \varepsilon, \nonumber
\end{align}
where the variables are defined as \( x_1 = \alpha_1^\prime \mathbf{z}, x_2 = \alpha_2^\prime \mathbf{z}, x_3 = \alpha_3^\prime \mathbf{z} \), with \( \mathbf{z} = \left( z_1, z_2, z_3 \right)^\prime \sim N(\mathbf{0}, \mathbf{I}_3) \) {and the noise $\varepsilon$, satisfying $E(\varepsilon)= 0$ and  
$E(\varepsilon^2) < \infty$, is independent of $\mathbf{z}$.} 
The coefficients are $ \alpha_1 = (1, 0, 0)^\prime, \alpha_2 = \left( \frac{1}{\sqrt{1+b^{2}}}, \frac{b}{\sqrt{1+b^{2}}}, 0 \right)^\prime$, and 
$\alpha_3 =$ $\left( \frac{1}{\sqrt{2+100b^2}}, \frac{10b}{\sqrt{2+100b^2}}, \frac{1}{\sqrt{2+100b^2}} \right)^\prime$, with \( b > 0 \), and \( \beta > 1 \). Thus, we have {\(E(x_i) = 0\) and \(E(x_i^2) =  1\), for 
$i=1,2,3.$ } 
\end{example}

We evaluate OGA's performance in this constructed example, focusing on its failure to correctly select the relevant variable \( x_{2} \), as illustrated in Table~\ref{tab:example1_OGA} below.

\begin{table}[H]
\centering
\caption{Variable selection performance of the OGA in Example 1.}
\label{tab:example1_OGA}
\begin{tabular}{cccc}
  \toprule
  \multicolumn{2}{c}{\textbf{Iteration 1}} & \multicolumn{2}{c}{\textbf{Iteration 2}} \\ 
  \cmidrule(lr){1-2} \cmidrule(lr){3-4} 
  \( \left| \mu_{J_{1}, 1} \right| \) & \( \beta + \frac{1}{\sqrt{1+b^{2}}} \)   & \( \left| \mu_{J_{2}, 1} \right| \)  & \( 0 \)  \\
  \( \left| \mu_{J_{1}, 2} \right| \) & \( \frac{\beta}{\sqrt{1+b^{2}}} + 1 \)   & \( \left| \mu_{J_{2}, 2} \right| \)    & \(\frac{b^{2}}{1+b^{2}} \)  \\
  \( \left| \mu_{J_{1}, 3} \right| \) & \( \frac{\beta}{\sqrt{2+100b^{2}}} 
    + \frac{1+10b^{2}}{\sqrt{\left( 1+b^{2} \right)\left( 2+100b^{2} \right)}} \)    & \( \left| \mu_{J_{2}, 3} \right| \)  & \(\frac{10b^{2}}{\sqrt{\left( 1+b^{2} \right) \left( 2+100b^{2} \right)}} \)  \\
  \bottomrule
\end{tabular}
\end{table}

In iteration 1 of Table~\ref{tab:example1_OGA},  \( \left| \mu_{J_1, 1} \right| \) is the largest, allowing OGA to correctly select \( x_1 \). However, in iteration 2, \( \left| \mu_{J_2, 3} \right| \) becomes the largest, causing OGA to incorrectly select \( x_3 \) instead of the true variable \( x_2 \). Therefore, this example shows that the marginal-correlation criterion used by OGA may incorrectly evaluate variable importance in the presence of strong correlations among predictors. As a result, OGA may fail to recover relevant variables even in relatively simple settings, since marginal correlations do not necessarily reflect the true incremental contribution of a predictor once other variables are present.

\subsection{Gram-Schmidt Forward Regression}
Section \ref{sec2.2} highlights a key limitation of the OGA selection criterion: its reliance on marginal contributions may lead to incorrect variable selection. In ultra-high dimensional settings, where predictors are often strongly correlated, effective variable selection requires a criterion that can better account for such dependence. To address this issue, we consider a Gram–Schmidt implementation of forward regression, denoted by GSFR, which replaces marginal ranking with orthogonalized ranking and highlights the unique contribution of each candidate predictor conditional on the variables already selected. Revisiting Example 1, we show that GSFR achieves more accurate variable selection than OGA, particularly when the predictors are highly correlated. 

The GSFR employs an iterative procedure to evaluate the unique contribution of each variable to the residuals. At each step, the variable with the largest unique contribution is selected. Specifically, the selection criteria are given by

\begin{equation}
j_{k+1} = \arg \max_{1 \leq j \leq p} |\mu_{J_k, j}|, \quad
\hat{j}_{k+1} = \arg \max_{1 \leq j \leq p} |\hat{\mu}_{\hat{J}_k, j}|.
\label{criterion}
\nonumber
\end{equation}

Analogous to correlations, for \(i \notin J\), the unique contribution measure is defined as follows:

\begin{equation}
\begin{aligned}
{\mu_{J, i}} &= 
\frac{
E\left[\left\{y(\mathbf{x}) - y_{J}(\mathbf{x})\right\} x_{i;J}^{\bot}\right]
}{
E\left[{\left ( y(\mathbf{x}) - y_{J}(\mathbf{x}) \right ) }^{2}\right]^{1/2}
E\left[{x_{i;J}^{\bot}}^{2}\right]^{1/2}
}, \\
{\hat{\mu}_{J, i}} &= 
\frac{
n^{-1} \sum_{t=1}^{n} \left(y_{t} - \hat{y}_{t; J}\right) \hat{x}_{t i; J}^{\bot}
}{
\left(n^{-1} \sum_{t=1}^{n} \left(y_{t} - \hat{y}_{t; J}\right)^{2}\right)^{1/2}
\left(n^{-1} \sum_{t=1}^{n} \left(\hat{x}^{\bot}_{t i; J}\right)^{2}\right)^{1/2}
}.
\end{aligned}
\label{correlation}
\nonumber
\end{equation}

Here, \(x_{i;J}^{\bot} = x_{i} - x_{i}^{(J)}\) denotes the orthogonal complement of \(x_{i}\) with respect to the linear space spanned by \(\{x_{j} : j \in J\}\), where \(x_{i}^{(J)}\) is its projection onto that space. Similarly, \(\hat{x}_{t i; J}^{\bot} = x_{ti} - \hat{x}_{t i; J}\) represents the corresponding sample version. For \(i \in J\), we naturally define {\(\mu_{J, i} = 0\) and \(\hat{\mu}_{J, i} = 0\).}

GSFR applies the Gram-Schmidt orthogonalization process to iteratively update the candidate set by computing the orthogonal complement of each unselected variable with respect to the selected ones. This ensures that the updated candidate set contains variables orthogonal to the selected variables, thereby capturing their unique contributions. Unlike (\ref{ref1}) in OGA, which evaluates marginal contributions based on the original variables, GSFR focuses on unique contributions, effectively addressing the challenges posed by high correlations among variables. To simplify the computations, the following proxies are used ({for simplicity, we kept the same notation}):

\begin{equation}
\mu_{J, i} = \frac{E\left[\left\{y(\mathbf{x}) - y_{J}(\mathbf{x})\right\} x_{i;J}^{\bot}\right]}{E\left[{x_{i;J}^{\bot}}^{2}\right]^{1/2}}, \quad 
\hat{\mu}_{J, i} = \frac{n^{-1} \sum_{t=1}^{n} \left(y_{t} - \hat{y}_{t; J}\right) \hat{x}_{t i; J}^{\bot}}{\left(n^{-1} \sum_{t=1}^{n} \left(\hat{x}^{\bot}_{t i; J}\right)^{2}\right)^{1/2}+\rho_{1}\left ( \frac{\log p_{n} }{n}  \right ) ^{1/2}}.
\label{ref2}
\end{equation}

The above formulation indicates  that the population version of GSFR corresponds to PrWGA in \citet{borodinProjectionGreedyAlgorithm2021}. Additionally, the term \( \rho_{1} \left( \frac{\log p_{n}}{n} \right)^{1/2} \), where \( \rho_{1} \geq 0 \), is an adjustment introduced to enhance numerical stability, ensuring that the selection criterion avoids selecting singular variable sets. Notably, \( \rho_{1} \) can be chosen arbitrarily small with a theoretical guarantee. Therefore, we recommend that practitioners always include a small value, such as \( 1 \times 10^{-6} \), as this does not affect the reliability of the algorithm’s selection path.

For \(i \notin J\), the following equivalence holds:  
\begin{equation}
E\left[\left\{y(\mathbf{x}) - y_{J}(\mathbf{x})\right\}x_{i;J}^{\bot}\right] = E\left[\left\{y(\mathbf{x}) - y_{J}(\mathbf{x})\right\}x_{i}\right], \quad 
\sum_{t=1}^{n}\left(y_{t} - \hat{y}_{t; J}\right)\hat{x}_{t i; J}^{\bot} = \sum_{t=1}^{n}\left(y_{t} - \hat{y}_{t; J}\right)x_{t i}.
\nonumber
\end{equation}  
Thus, (\ref{ref2}) can be expressed as:  
\begin{equation}
\mu_{J, i} = \frac{E\left[\left\{y(\mathbf{x}) - y_{J}(\mathbf{x})\right\} x_{i}\right]}{E\left[{x_{i;J}^{\bot}}^{2}\right]^{1/2}}, \quad 
\hat{\mu}_{J, i} = \frac{n^{-1} \sum_{t=1}^{n} \left(y_{t} - \hat{y}_{t; J}\right) x_{t i}}{\left(n^{-1} \sum_{t=1}^{n} \left(\hat{x}^{\bot}_{t i; J}\right)^{2}\right)^{1/2}+\rho_{1}\left ( \frac{\log p_{n} }{n}  \right ) ^{1/2}}.
\label{ref3}
\end{equation}  

In comparison, (\ref{ref1}) and (\ref{ref3}) differ only in their denominators, which  affects the normalization of variables. This distinction explains the key difference between the selection criteria of GSFR and OGA. In the {GSFR} selection criterion, only the unique part of a variable, unexplained by the previously selected variables, contributes to the selection process. OGA may incorrectly normalize this unique part, thereby impacting the selection outcome. After one variable is selected, both methods perform component-wise linear regression with orthogonalized variables to compute the residuals, thereby reducing computational cost.

\par
We now illustrate the advantage of the proposed GSFR over OGA by applying GSFR to Example 1, where the predictors are highly correlated. The specific performance of GSFR is summarized in the Table~\ref{tab:example1_GSReg} below.

\begin{table}[H]
\centering
\caption{Performance of GSFR in Example 1.}
\label{tab:example1_GSReg}
\begin{tabular}{cccc}
  \toprule
  \multicolumn{2}{c}{\textbf{Iteration 1}} & \multicolumn{2}{c}{\textbf{Iteration 2}} \\ 
  \cmidrule(lr){1-2} \cmidrule(lr){3-4} 
  \( \left| \mu_{J_{1}, 1} \right| \) & \( \beta + \frac{1}{\sqrt{1+b^{2}}} \)   & \( \left| \mu_{J_{2}, 1} \right| \)  & \( 0 \)  \\
  \( \left| \mu_{J_{1}, 2} \right| \) & \( \frac{\beta}{\sqrt{1+b^{2}}} + 1 \)   & \( \left| \mu_{J_{2}, 2} \right| \)    & \(\frac{b}{\sqrt{1+b^{2}} } \)  \\
  \( \left| \mu_{J_{1}, 3} \right| \) & \( \frac{\beta}{\sqrt{2+100b^{2}}} 
    + \frac{1+10b^{2}}{\sqrt{\left( 1+b^{2} \right)\left( 2+100b^{2} \right)}} \)    & \( \left| \mu_{J_{2}, 3} \right| \)  & \(\frac{10b^{2}}{ \sqrt{\left ( 1+b^{2} \right ) \left ( 1+100b^{2} \right ) } } \)  \\
  \bottomrule
\end{tabular}
\end{table}
In this example, we observe that \(\left| \mu_{J_{2}, 2} \right| > \left| \mu_{J_{2}, 3} \right|\), indicating that  GSFR correctly selects \(x_{2}\) based on its unique contribution, thereby improving the accuracy of variable selection. This improvement occurs because GSFR correctly normalizes the unique part of each variable. In contrast, OGA fails to normalize correctly, leading to an overestimation of the unique contribution of \(x_{3}\) to the response.

Conceptually, GSFR implements forward regression through a Gram–Schmidt orthogonalization scheme. At each step, candidate predictors are orthogonalized with respect to the variables already selected, and their contributions are evaluated through correlations with the resulting orthogonal components. This formulation leads to a selection rule based on the unique contribution of each predictor after removing the effects of previously selected variables.

The proposed approach is related to the projection-based forward regression or forward selection  procedure in \citet{wang2009forward}, which also aim to reduce the residual sum of squares when augmenting the model sequentially. However, GSFR differs  in how candidate variables are evaluated along the selection path. In particular, PBFR relies on repeated projection operations involving augmented design matrices, whereas GSFR evaluates predictors through correlations with orthogonalized variables obtained via the Gram–Schmidt procedure. This formulation avoids repeated projection steps and leads to a computationally efficient and numerically stable algorithm in ultra-high dimensional settings. Moreover, GSFR remains well-defined even in situations where projection-based procedures may encounter singularity issues, as illustrated in the following example.

\begin{example}
{Consider} the model,
    $y = 2 x_{1} + x_{2} + \varepsilon$, 
where the variables are defined {in the same way as those of Example 1, but the}
coefficients are {changed to } 
\( \alpha_1 = (2, 0,0)^\prime, \alpha_2 = \left( 0, 1,0 \right)^\prime\), and  \(\alpha_3 = \left( 1, 0, \eta \right)^\prime \), with a small constant \( \eta \ge 0 \).
\end{example}

In this example, it is easy to verify that GSFR can still correctly select \(x_1\) and \(x_2\), even though \(x_1\) and \(x_3\) are nearly collinear as  \(\eta\) approaches 0. In such cases, the PBFR may become numerically problematic when the augmented design matrix is singular or nearly singular. In contrast, the Gram–Schmidt formulation remains well defined in this setting.



 \subsection{A Path-Based Model Size Selection Rule}

In this subsection, we propose a new path-based model size selection rule. The optimal model size is defined as the number of variables selected when the last relevant predictor enters the model along the forward selection path. The proposed estimator consistently identifies this point by detecting a pronounced decline in the incremental explanatory power of newly selected variables.

The intuition is straightforward. Under a sparse model, each relevant variable typically produces a substantial reduction in the residual variance when it enters the model. Once all relevant variables have been selected, the remaining predictors primarily capture noise and therefore yield only negligible further reductions. As a result, the sequence of incremental variance reductions is expected to display a clear break. The proposed ratio criterion is designed to detect this change and thereby identify the appropriate stopping point. Since GSFR generates an orthogonalized forward path, model size selection can be based naturally on the successive declines in orthogonalized incremental contributions along that path.
 


We {define} the optimal model size along a GSFR selection path by  
\begin{equation}
\tilde{k}_n = \min\{k : 1 \leq k \leq K_n, N_n \subseteq \hat{J}_k\} \quad (
\mbox{with}\quad \min \emptyset = K_n),  
\label{ref20}
\nonumber
\end{equation}  
where \(N_n\) {is} the number of relevant variables in the model 
{and} \( K_n = O((n / \log p_n)^{1/2}) \) represents the total number of iterations of GSFR. {When the selection path does not contain $N_n$, we define $\tilde{k}_n =\min \emptyset = K_n$.}

Next, we define the population and sample versions of the ratio criterion based on the differenced sequence of the {residual variances} along the GSFR selection path. They are given by
\begin{equation}
\triangle_{\hat{J}_m} =
\begin{cases}
\frac{\sqrt{\sigma_{\hat{J}_m}^2 - \sigma_{\hat{J}_{m+1}}^2}}{\sqrt{\sigma_{\hat{J}_{m-1}}^2 - \sigma_{\hat{J}_m}^2}}, & \text{if } m \leq \tilde{k}_n, \\
1, & \text{if } m > \tilde{k}_n,
\end{cases}
\quad
\hat{\triangle}_{\hat{J}_m} = \frac{\sqrt{\hat{\sigma}_{\hat{J}_m}^2 - \hat{\sigma}_{\hat{J}_{m+1}}^2} + \rho_{2}\frac{1}{n^{1.5\gamma + \varepsilon_0}}}{\sqrt{\hat{\sigma}_{\hat{J}_{m-1}}^2 - \hat{\sigma}_{\hat{J}_m}^2} + \rho_{2}\frac{1}{n^{1.5\gamma + \varepsilon_0}}},
\label{ref21_22}
\nonumber
\end{equation}  
where \(\sigma_J^2 = E \left\{ \left( y(\mathbf{x}) - y_{J}(\mathbf{x}) \right)^2 \right\}\) and \(\hat{\sigma}_J^2 = n^{-1} \sum_{t=1}^n \left(y_t - \hat{y}_{t;J}\right)^2\) denote the population and sample versions of the mean squared error when the target variable is regressed on the variables in \(J\), respectively. The adjustment term \(\rho_{2}\frac{1}{n^{1.5\gamma + \varepsilon_0}}\), where \(\gamma\) and \(\varepsilon_0\) are defined in Assumption~\ref{asm6} and \(\rho_{2} > 0\), is introduced to enhance numerical stability in {empirical analysis}. Notably, \(\rho_{2}\) can be set arbitrarily small in practice, and its choice does not affect the theoretical performance of the proposed model size selection criterion.


Through simplification, we obtain  
\begin{equation}
\triangle_{\hat{J}_m} = \frac{\left| \mu_{\hat{J}_{m}, \hat{j}_{m+1}} \right|}{\left| \mu_{\hat{J}_{m-1}, \hat{j}_{m}} \right|}, \quad 
\hat{\triangle}_{\hat{J}_m} = \frac{\left| \hat{\mu}_{\hat{J}_{m}, \hat{j}_{m+1}} \right| + \rho_2 \frac{1}{n^{1.5\gamma + \varepsilon_0}}}{\left| \hat{\mu}_{\hat{J}_{m-1}, \hat{j}_{m}} \right| + \rho_2 \frac{1}{n^{1.5\gamma + \varepsilon_0}}},
\label{ref24}
\nonumber
\end{equation}
with \(\rho_1 = 0\) in this setting. This ratio effectively captures significant changes in the unique contributions of consecutive variables.

Building on the definition of \(\hat{\triangle}_{\hat{J}_m}\), we propose a data-driven criterion for selecting the model size. The estimated model size \(\hat{k}_n\) is defined as  
\begin{equation}
\hat{k}_n = \arg\min_{1 \leq m \leq K_n -1} \hat{\triangle}_{\hat{J}_m},
\label{ref23}
\nonumber
\end{equation}
which serves as a consistent estimator of the optimal model size \(\tilde{k}_n\).



The proposed rule is computationally convenient because it uses quantities already computed along the forward path and incurs negligible additional cost. Relative to information-criterion-based or cross-validation-based alternatives, it requires little additional tuning and is straightforward to implement in ultra-high dimensional settings.

\subsection{Implementation  of GSFR }\label{imple}
In this subsection, we present the implementation procedure of the GSFR algorithm, which consists of the following steps:

{\itshape
\begin{enumerate}[label=(\arabic*)]
    \item 
    \textbf{Initialization:} Set $\hat{J}_0 = \emptyset$, the maximum number of iterations $m_{\mathrm{stop}} = K_n$, and the iteration index $m = 1$.

    \item 
    \textbf{Variable Selection:} At the $m${\text{th}} step, for each predictor variable $j = 1, \dots, p$, compute $\hat{\mu}_{\hat{J}_m-1,j}$, and select
    \[
    \hat{j}_{m} = \arg\max_{1 \le j \le p} \left|\hat{\mu}_{\hat{J}_m-1,j}\right|.
    \]

    \item 
    \textbf{Prediction and Residuals:} Update predictions and compute residuals as
    \[
    \hat{y}_{m}(\mathbf{x}_t) = \hat{y}_{m-1}(\mathbf{x}_t) + \hat{\beta}_{\hat{j}_{m}} \hat{x}_{t \hat{j}_{m}; \hat{J}_m-1}^{\perp}, \quad
    U_t^{(m)} = y_t - \hat{y}_{m}(\mathbf{x}_t),
    \]
    where
    \[
    \hat{\beta}_{\hat{j}_{m}} = \frac{\sum_{t=1}^{n}y_t\hat{x}_{t \hat{j}_{m}; \hat{J}_{m-1}}^{\perp}}{\sum_{t=1}^{n}(\hat{x}_{t \hat{j}_{m}; \hat{J}_{m-1}}^{\perp})^2}, \quad (1 \le t \le n).
    \]

    \item 
    \textbf{Orthogonalization:} Update the orthogonalized predictor variables. For \( j = 1, \dots, p \), compute
    \[
    \hat{x}_{t j; \hat{J}_{m}} = \hat{x}_{t j; \hat{J}_{m-1}} + \hat{\alpha}_{j ; \hat{j}_{m}} \hat{x}_{t \hat{j}_{m}; \hat{J}_{m-1}}^{\perp}, \quad
    \hat{x}_{t j; \hat{J}_{m}}^{\perp} = x_{t j} - \hat{x}_{t j; \hat{J}_{m}},
    \]
    where
    \[
    \hat{\alpha}_{j ; \hat{j}_{m}} = \frac{\sum_{t=1}^{n} x_{t j} \hat{x}_{t \hat{j}_{m}; \hat{J}_{m-1}}^{\perp}}{\sum_{t=1}^{n} (\hat{x}_{t \hat{j}_{m}; \hat{J}_{m-1}}^{\perp})^2}, \quad (1 \le t \le n).
    \]

    \item 
    \textbf{Iteration Check:} Increment \(m\) by one. If \(m \le m_{\mathrm{stop}}\), return to step (2); otherwise, proceed to the next step.

    \item 
    \textbf{Model Size Selection:} Compute
    \[
    \hat{k}_n = \arg\min_{1 \leq m \leq K_n-1} \hat{\triangle}_{\hat{J}_m},
    \]
    and select the estimated optimal model with variable index set \( \hat{J}_{\hat{k}_n} \).
\end{enumerate}
}

The GSFR algorithm is computationally efficient and numerically robust, allowing for fast implementation even in ultra-high dimensional settings. When two variables are strongly correlated, the selection of one substantially reduces the orthogonalized incremental contribution of the other, making it unlikely to be selected in subsequent steps. This mechanism also helps avoid singularity problems. In addition, GSFR tends to produce parsimonious models, thereby reducing the risk of overfitting.

Table~\ref{tab:method_comparison} summarizes the main differences among GSFR, PBFR, and OGA. PBFR repeatedly augments the current model and selects the predictor that produces the largest reduction in residual variance, making it increasingly costly as the model grows. In contrast, GSFR has complexity $O(npK)$, comparable to OGA and lower than the $O(npK^3)$ complexity of PBFR in \cite{wang2009forward}. In addition, PBFR may become numerically unstable under strong collinearity, whereas OGA may misidentify relevant predictors because marginal correlations can be misleading in highly correlated designs.

\begin{table}[htbp!]
\centering
\caption{Comparison of GSFR with related greedy variable selection methods}
\label{tab:method_comparison}
\renewcommand{\arraystretch}{1.35}
\begin{tabular}{p{1.5cm} p{3.2cm} p{2.5cm} p{6.3cm} p{2.2cm}}
\hline
\textbf{Method} & \textbf{\makecell[t]{Selection\\Criterion}} & \textbf{\makecell[t]{Model Size\\Selection}} & \textbf{\makecell[t]{Behavior in\\Ultra-High Dimensions}} & \textbf{Complexity} \\
\hline
PBFR & RSS reduction via augmented projections & BIC & Repeated projection updates impose increasing computational burden as the model grows and may become numerically unstable when predictors are highly collinear & $O(npK^3)$ \\
OGA & Marginal correlation with residuals & \makecell[l]{HDIC /\\ HDAIC} & May mis-rank relevant variables because marginal correlations can be distorted by strong collinearity & $O(npK)$ \\
GSFR & Correlation with orthogonalized predictors & Variance-ratio stopping rule & Recursive orthogonalization reduces the computational burden of sequential updates and improves stability in strongly collinear settings & $O(npK)$ \\
\hline
\end{tabular}

\vspace{0.2cm}
\parbox{0.95\linewidth}{\footnotesize \textit{Note:} $K$ denotes the maximum number of iterations. For GSFR, $K = O\!\left(\sqrt{n/\log p}\right)$.}
\end{table}

\section{Theoretical Properties}\label{sec3}

In this section, we establish the asymptotic properties of the GSFR algorithm {when} 
both the sample size $n$ and the dimensionality $p$ diverge to infinity. 
We begin by introducing a set of regularity assumptions. The detailed 
proofs are provided in the supplementary material.

 

\begin{assumption}
{The} dimensionality \( p_n \to \infty \) and satisfies \( \log p_n = o(n) \), 
{where the subscript $n$ is added to signify its dependence on $n$}.
\label{assumption1}
\end{assumption} 

\begin{assumption}
The error terms \(\{\varepsilon_t\}\) {satify}
\[
E\{\exp(s \varepsilon_t)\} < \infty, \quad \text{for } |s| \leq s_0,
\]
where \( s_0 > 0 \) is a constant.
\label{assumption2}
\end{assumption}

\begin{assumption}
There exists a constant \(s_1 > 0\) such that:
\[
\limsup_{n \to \infty} \max_{1 \leq j \leq p_n} E\left\{\exp\left(s_1 z_j^2\right)\right\} < \infty,
\]
where {\( z_j = \frac{x_j}{\sigma_j} \) with \( \sigma_j^2 = E(x_j^2) \)}.
\label{assumption3}
\end{assumption}

\begin{assumption}
{The weak sparsity condition holds}:
\begin{equation}
\sup_{n \geq 1} \sum_{j=1}^{p_n} \left|\beta_j \sigma_j\right| < \infty.
\label{assumption4}	
\end{equation}
\end{assumption}

\begin{assumption}
Let \(K_n\) denote the prescribed upper bound on the number of GSFR iterations. Define:
\begin{equation}
\boldsymbol{\Gamma}(J) = E\{\mathbf{z}(J) \mathbf{z}^\prime(J)\}, \quad \mathbf{g}_i(J) = E\{z_i \mathbf{z}(J)\},
\label{assumption5_equation1}
\end{equation}
where \(\mathbf{z}(J)\) is the subvector of \((z_1, \dots, z_p)^\prime\) given by the subset \(J 
\subset \{1,\ldots, p\} \). We assume that for some \(\delta > 0\) and \(M > 0\), and for all sufficiently large \(n\),
\begin{equation}
\min_{1 \leq \sharp(J) \leq K_n}\lambda_{min}(\boldsymbol{\Gamma}(J)) > \delta, 
\label{assumption5_equation2}
\end{equation}
\begin{equation}
\max_{1 \leq \sharp(J) \leq K_n, i \notin J} \left\|\boldsymbol{\Gamma}^{-1}(J) \mathbf{g}_i(J)\right\|_1 < M,
\label{assumption5_equation3}
\end{equation}
where \(\sharp(J)\) is the cardinality of \(J\) {and $\lambda_{min}(\boldsymbol{\Gamma})$ 
denotes the minimum eigenvalue of the matrix $\boldsymbol{\Gamma}$}.
\end{assumption}

Assumption 1 allows the predictor dimension to diverge to infinity at an exponentially fast rate, implying that the predictor dimension can be substantially larger than the sample size \(n\), as discussed in \citet{fan2008sure}. Assumptions 2 and 3 ensure that the noise and predictors do not exhibit excessively heavy tails. These conditions facilitate the derivation of exponential bounds on moderate deviation probabilities essential for the analysis, including those for the sample correlation matrix of \(\mathbf{x}_t\). Assumption 4 allows for small but nonzero coefficients, reflecting a flexible sparsity structure where the minimum coefficients may shrink as the sample size increases. Assumption 5 guarantees that the \( p_n \)  feature variables are linearly independent in the Hilbert space, ensuring that the Gram-Schmidt orthogonalization procedure is well-defined. It also guarantees the invertibility of the correlation matrix \( \boldsymbol{\Gamma}(J) \), with \( 1 \le \sharp(J) \leq K_n \). This assumption is not inconsistent with our discussion of how GSFR alleviates singularity concerns in practice. Indeed, when two variables are strongly correlated, the selection of one substantially reduces the orthogonalized incremental contribution of the other, making the latter unlikely to be selected in subsequent steps. Consequently, the theoretical covariance conditions do not need to accommodate nearly redundant variables as simultaneously active contributors along the selected path. Furthermore, the second condition of Assumption 5 bounds the \( L_1 \)-norm of the regression coefficients for \( x_i \) on the subset \( \mathbf{x}(J) \). Assumption 5 holds even in the presence of strong correlations among predictors, as demonstrated in \citet{ing2020model}.

The following theorem establishes the convergence rate of the GSFR algorithm.
\begin{theorem}
	Assume that Assumptions 1 to 5 hold. Suppose  $K_{n} \rightarrow \infty$  such that $K_{n}=   O\left(\left(n / \log p_{n}\right)^{1 / 2}\right)$. Then for GSFR,
	
\begin{equation}
	\max _{1 \leq m \leq K_{n}}\left(\frac{E\left[\left\{y(\mathbf{x})-\hat{y}_{m}(\mathbf{x})\right\}^{2} \mid y_{1}, \mathbf{x}_{1}, \ldots, y_{n}, \mathbf{x}_{n}\right]}{m^{-1}+n^{-1} m \log p_{n}}\right)=O_{p}(1) .
\nonumber
\end{equation}
\label{theorem1}
\end{theorem}

\begin{remark}
Theorem \ref{theorem1} establishes a uniform convergence rate for GSFR, showing that the conditional mean squared error decreases at a rate of \( O_{p}(m^{-1} + n^{-1}m \log p_n) \). Our results demonstrate that GSFR attains a convergence rate comparable to OGA under weak sparsity, providing additional theoretical insight into the behavior of related forward selection procedures such as those in \citet{wang2009forward}, if they stop before \( K_n \). The standard bias-variance tradeoff indicates that \( m \) should not exceed \( O((n / \log p_n)^{1/2}) \) and that setting \( m = K_n \) achieves the optimal balance. The term \( m^{-1} \) in our convergence rate is consistent with the theoretical results of \citet{borodinProjectionGreedyAlgorithm2021}, arguing that further improvement beyond \( m^{-1} \) is not achievable under weak sparsity. This result underscores the competitiveness of our method and confirms the effectiveness of GSFR in ultra-high dimensional settings.
\end{remark}

Next, we establish the sure screening property of GSFR under strong sparsity, where irrelevant variables have zero coefficients, and the number of relevant variables with nonzero coefficients grows at a controlled rate with \( n \). 

\begin{assumption}\label{asm6}
We assume the following conditions hold:
\begin{enumerate}[label=(\alph*)]
    \item There exists a sufficiently small \(\varepsilon_0 > 0\) and a constant \(0 < \gamma < \frac{2(1 - \varepsilon_0)}{3}\) such that  
    \begin{equation}
        n^{1.5\gamma+\varepsilon_0} = o\left(\left(\frac{n}{\log p_n}\right)^{1/2}\right).
        \label{assumption6_equation1}
    \end{equation}
    \item The nonzero coefficients satisfy  
    \begin{equation}
        \liminf_{n \to \infty} n^\gamma \min_{1 \leq j \leq p_n : \beta_j \neq 0} \beta_j^2 \sigma_j^2 > 0.
        \label{assumption6_equation2}
    \end{equation}
\end{enumerate}
\end{assumption}

Assumption \ref{asm6}, a stronger condition than weak sparsity and widely adopted in the ultra-high-dimensional variable selection literature, is essential to ensure the sure screening property of GSFR. Specifically, {Condition} (\ref{assumption6_equation1}) provides a lower bound on the growth rate of \( K_n \), the number of iterations needed to achieve the convergence rate in Theorem \ref{theorem1}. Meanwhile, 
{Condition} (\ref{assumption6_equation2}) constrains the decay rate of the smallest nonzero coefficient to at most \( O\left(n^{-\gamma/2}\right) \). Intuitively, if any nonzero coefficient decays too rapidly, it cannot be consistently identified. Together with the weak sparsity assumption, these conditions ensure that the number of relevant variables, denoted by \( \sharp(N_n) \), grows at a controlled rate of \( O\left(n^{\gamma/2}\right) \) as \( n \) increases, allowing GSFR to recover the true model structure effectively.

\begin{theorem}
Assume that Assumptions \ref{assumption1} to \ref{asm6} {hold}. Let \( N_n = \{1 \leq j \leq p_n : \beta_j \neq 0 \} \) denote the set of relevant input variables. Suppose \( K_n / n^{\gamma} \to \infty \) and \( K_n = O\left(\left(n / \log p_n\right)^{1/2}\right) \). Then, 
\[
\lim_{n \to \infty} P\left(N_n \subset \hat{J}_{K_n}\right) = 1,
\]
where \( \hat{J}_{K_n} \) is the set of variables selected by GSFR after \( K_n \) iterations.
\label{theorem2}
\end{theorem}

\begin{remark}
Theorem \ref{theorem2} demonstrates that, with probability approaching one, the GSFR algorithm identifies all relevant predictors within \( K_n = O\left(\left(n / \log p_n\right)^{1/2}\right) \) iterations. This ensures that the number of iterations required by GSFR remains smaller than the sample size \( n \), effectively reducing computational cost. In contrast, the standard FR algorithm such as that in \citet{wang2009forward} requires \( n \) iterations. In practice, the final model size is determined by the path-based stopping rule, which helps exclude redundant variables and typically yields a parsimonious selected model.


\nonumber
\end{remark}

We now establish the variable selection consistency of GSFR using the model size selection criterion proposed in Section 2.

\begin{theorem}
Under the same notation and assumptions as {those} in Theorem \ref{theorem2}, suppose \( K_n / n^{\gamma} \to \infty \) and \( K_n = O((n / \log p_n)^{1/2}) \). Then
\begin{equation}
\lim_{n \to \infty} P(\hat{k}_n = \tilde{k}_n) = 1.
\nonumber
\end{equation}
\label{theorem3}
\end{theorem}

\begin{remark}

Theorem 3 establishes that the proposed stopping rule consistently identifies the optimal model size along the GSFR path. This provides theoretical support for using GSFR as a data-driven screening procedure that combines a truncated forward path with consistent model size determination. The result is particularly useful in ultra-high dimensional settings, where stable screening and parsimonious model selection are both important. Furthermore, relative to the just-in-time stopping approach in \citet{JMLR:v25:23-0882}, our method terminates at a sufficiently small \(K_n\), balancing convergence rate and computational efficiency, and provides a stronger screening property.


\end{remark}

\section{Numerical Properties}\label{sec4}
In this section, we use simulation and real data analysis to assess the performance of the proposed {GSFR} procedure in finite samples.

\subsection{Simulation}
 We conduct Monte Carlo experiments on several data generating models to evaluate the effectiveness of GSFR and to compare it with some existing approaches. For greedy algorithms, we take OGA with the HDBIC rule in \citet{ingStepwiseRegressionMethod2011}, PBFR with the BIC rule in \citet{wang2009forward}, and GSFR with \(n\) iterations under our proposed model size selection criterion (GSFRn) as benchmarks. For penalized methods, Lasso, Adaptive Lasso, SIS-SCAD, ISIS-SCAD, and LARS are included in the comparison. The {maximum number of 
iterations} \( K_n \) is set to \( \left\lfloor 5 \left( \frac{n}{\log p_n} \right)^{1/2} \right\rfloor \), as suggested in \citet{ingStepwiseRegressionMethod2011}. The adjustment terms are set to \(1 \times 10^{-6}\). For every setup, we conduct \(T = 1000\) simulation replications.

 To extensively study the variable selection performance, we evaluate several indices. Let \( \hat{J}_{\hat{k}_n}^{(t)} \) denote the selected model in the \( t \)-th simulation replication and, 
 {again,} \( N_n \) represents the true model. The Coverage Probability is calculated as \( T^{-1} \sum_{t} I\left(\hat{J}_{\hat{k}_n}^{(t)} \supset N_n\right) \),  measuring the probability that all relevant variables are selected. We also compute the Percentage of False Negatives as \( T^{-1} \sum_{t} \sharp(N_n \setminus \hat{J}_{\hat{k}_n}^{(t)}) /\sharp(N_n) \), which quantifies the average proportion of relevant variables in \( N_n \) that are not selected by the method. Similarly, the Percentage of False Positives is given by \( T^{-1} \sum_{t} \sharp(  \hat{J}_{\hat{k}_n}^{(t)} \setminus N_n) /\sharp(N_n^c )  \), reflecting the average proportion of irrelevant variables incorrectly selected. To assess model size selection accuracy, we track the Best Model Size for greedy methods in simulation replications where all relevant variables are successfully selected. This represents the model size at the point when all relevant variables are first included in the selection path. A more accurate method should yield a smaller Best Model Size while maintaining a high Coverage Probability. Additionally, we record the Selected Model Size in the same replications. A method with a Selected Model Size closer to the Best Model Size is considered more precise. Finally, to assess both in-sample and out-of-sample performance, we report the average RSS of the selected model and the MSPE for predicting \( y_{n+1} \).

Consider first Example 3, which examines the performance of GSFR relative to greedy alternatives. In this example, the relationship among variables follows a block-wise compound symmetry correlation structure, predictors within the same block exhibit the same degree of correlation with one another.

\noindent{\bf Example 3}. 
We consider the following linear model:  
\begin{align}
     y_{t} = \sum_{g=1}^{G} \sum_{j=1}^{q} \beta^{g}_{j} x^{g}_{t j} +\varepsilon_{t},\quad  t = 1, \ldots, n, \quad
    x^{g}_{t j} = \theta w^{g}_{t} + d^{g}_{t j}, \nonumber
\end{align}  
where the predictor components \( (d^{g}_{t1}, \ldots, d^{g}_{tq}, w^{g}_{t})^{\prime} \), for \(1 \le g \le G\) and \(1 \le t \le n\), are i.i.d. normal with mean zero and covariance matrix \( \mathrm{I}_{p+G} \), with \( p = qG \). The noise term \( \varepsilon_t \) follows an i.i.d. standard normal distribution \( \mathrm{N}(0,1) \) and is independent of the predictors. In this model, the variables are correlated only within the same group. We consider cases with G = 10: \( \theta = 2 \), \( 0.9 \) and \( (n,p)=(100,2000)\), \( (200,4000)\). Within this simulation framework, we specify the nonzero coefficients as
\( (\beta^{1}_1, \beta^{2}_1, \beta^{3}_1, \beta^{4}_1, \beta^{5}_1) = (3.0, -3.5, 4, -2.8, 3.2) \), with all remaining coefficients set to zero. Under this setting, for any \( J \subseteq\{1, \ldots, p\} \) and \( 1 \leq i \leq p \) with \( i \notin J \), we have \( \lambda_{\min }(\boldsymbol{\Gamma}(J))\ge\frac{1}{1+\theta^{2}}>0 \) and \( \left\|\boldsymbol{\Gamma}^{-1}(J) \mathbf{g}_{i}(J)\right\|_{1} \leq 10 \), ensuring that Conditions (\ref{assumption5_equation2}) and (\ref{assumption5_equation3}) are satisfied.

\begin{table}[!htbp]
\centering
\caption{Variable selection results using GSFR, PBFR, OGA, and GSFRn method in Example 3. 
{The results are based on 1000 replications.}}
\label{tab:Example3}
\begin{adjustbox}{max width=\linewidth}
\begin{tabular}{cccccccc}
 \toprule
\multirow{2}{*}{\textbf{Method}} & 
\multirow{2}{*}{\makecell{Coverage \\ probability (\%)}} & 
\multicolumn{2}{c}{\makecell{Percentage of}} & 
\multirow{2}{*}{\makecell{Best \\ model \\ size}} & 
\multirow{2}{*}{\makecell{Selected \\ model \\ size}} & 
\multirow{2}{*}{RSS} & 
\multirow{2}{*}{MSPE} \\
\cline{3-4}
& & \makecell{false \\ negatives (\%)} & \makecell{false \\ positives (\%)} & & & & \\
  \midrule
  \multicolumn{8}{c}{ \(\theta = 2\) \quad $(n, p)$ = (100, 2000)} \\ 
GSFR &100 & 0.0000& 0.0956& 6.91& 6.91&  91.70& 1.1594 \\
 PBFR &100 & 0.0000& 1.4777& 6.91& 34.48&  51.74& 1.8380 \\
 OGA &98.10 & 1.9000& 0.1089 & 7.26& 7.11&  217.87& 6.0811\\
 GSFRn &100 & 0.0000&0.0956 & 6.91& 6.91&  91.70& 1.1594\\
 & & & & & & &  \\
 \multicolumn{8}{c}{\(\theta = 2\) \quad $(n, p)$ = (200, 4000)} \\
 GSFR &100 & 0.0000& 0.0109& 5.44& 5.44&  193.79& 1.0365 \\
 PBFR &100 & 0.0000& 0.0112& 5.44& 5.45&  193.58& 1.0391 \\
 OGA &100 & 0.0000& 0.0113 & 5.45& 5.45&  193.76& 1.0379\\
 GSFRn &100 & 0.0000& 0.0109& 5.44& 5.44&  193.79& 1.0365 \\
 & & & & & & &  \\
 \multicolumn{8}{c}{\(\theta = 0.9\) \quad $(n, p)$= (100, 2000)} \\
GSFR &100 & 0.0000& 0.0033& 5.07& 5.07&  93.62& 1.1296 \\
 PBFR &100 & 0.0000& 1.5188&  5.07& 35.30&  48.03& 1.9096 \\
 OGA &100 & 0.0000& 0.0037 & 5.07& 5.07&  93.61& 1.1300\\
GSFRn &100 & 0.0000& 0.0033& 5.07& 5.07&  93.62& 1.1296 \\
 & & & & & & &  \\
 \multicolumn{8}{c}{\(\theta = 0.9\) \quad $(n, p)$ = (200, 4000)} \\
GSFR &100 & 0.0000& 0.0000& 5.00&  5.00&  194.23& 1.0448 \\
 PBFR &100 & 0.0000& 0.0002& 5.00&  5.01&  194.04& 1.0456 \\
 OGA &100 & 0.0000& 0.0000& 5.00&  5.00&  194.23& 1.0448 \\
 GSFRn &100 & 0.0000& 0.0000& 5.00&  5.00&  194.23& 1.0448 \\
  \bottomrule
\end{tabular}
\end{adjustbox}
\end{table}

    Table~\ref{tab:Example3} {reports} the variable selection results using GSFR, PBFR, OGA, and GSFRn. Across all settings, GSFR attains full coverage of the relevant variables. It also achieves the lowest out-of-sample MSPE while maintaining the smallest average model size. These results demonstrate that GSFR provides a more favorable balance in variable selection. Notably, in the setting $(n,p) = (100, 2000)$ with $\theta = 2$, the correlation among predictors within the same group is high and the sample size is small. OGA recovers the relevant variables less effectively, with a false positive rate of 1.9\%, and exhibits poorest performance in both in-sample RSS and out-of-sample MSPE. PBFR is also affected by the insufficient sample size. It tends to overfit, selecting a relatively large model with low in-sample RSS, while its out-of-sample MSPE remains higher than that of GSFR.
   Moreover, PBFR is consistently outperformed by GSFRn in both false positive rate and out-of-sample MSPE throughout all scenarios, showing that the model size selection rule of GSFR algorithm is more effective than BIC at eliminating redundant variables. GSFRn and GSFR perform consistently, indicating that stopping at $K_n$ substantially reduces unnecessary computations. As $n$ increases, both the false positive rate and the MSPE of GSFR decrease. Noting that the oracle MSPE equals 1, this behavior provides empirical support for Theorem~\ref{theorem1}. Moreover, the average selected model size remains closely aligned with the best model size, which is consistent with Theorem~\ref{theorem3}. The decreasing best model size further suggests that GSFR exhibits the potential to achieve the oracle property, selecting only the relevant variables.

In Example 3, predictors from different groups are independent. We now turn to a more challenging setting, featuring a more complex correlation structure, to further assess GSFR's performance relative to other greedy approaches.

\noindent{\bf Example 4.} 
Consider the following linear model:
\begin{align*}
y_t &= \sum_{j=1}^{q-k} \beta^{1}_{j} x^{1}_{tj}
      + \sum_{g=2}^{G} \sum_{j=1}^{q-k} \beta^{g}_{j} x^{g}_{tj} 
      + \sum_{j=q-k+1}^{q} \beta^{G}_{j} x^{G}_{tj} 
      + \varepsilon_t, \quad t = 1, \ldots, n, \\[2pt]
x^g_{tj} &=
\begin{cases}
\theta w^1_t + d^1_{tj}, & 1 \le j \le q-k, \ g=1, \\[2pt]
\theta w^{g-1}_t + \theta w^g_t + d^g_{tj}, & 1 \le j \le k, \ g \ge 2, \\[2pt]
\theta w^g_t + d^g_{tj}, & k+1 \le j \le q-k, \ g \ge 2, \\[2pt]
\theta w^G_t + d^G_{tj}, & q-k+1 \le j \le q, \ g=G,
\end{cases}
\end{align*}
where the predictor components \( (d^{g}_{t1}, \ldots, d^{g}_{tq}, w^{g}_{t})^{\prime} \), for \(1 \le g \le G\) and \(1 \le t \le n\), are i.i.d. normal with mean zero and covariance matrix \( \mathrm{I}_{p+G} \), with \( p = G(q-k)+k \). The noise term \( \varepsilon_t \) follows an i.i.d. standard normal distribution \( \mathrm{N}(0,1) \) and is independent of the predictors. In this case, each group shares components with its neighboring groups through a subset of variables that belong to both groups, resulting in both within-group and inter-group correlations. We examine the cases with $G = 10$: \( \theta = 2 \), \( 1 \) and $(q,k,n,p) = (200, 40, 100, 1640)$, $(400, 80, 200, 3280)$. Within this simulation framework, we specify the nonzero coefficients as
\((\beta^{2}_1, \beta^{3}_1, \beta^{4}_1, \beta^{5}_1,\beta^{6}_1)= (3.0, -3.5, 4, -2.8, 3.2) \), with all remaining coefficients set to zero. 

\begin{table}[!htbp]
\centering
\caption{Variable selection results using GSFR, PBFR, OGA, and GSFRn method in Example 4. 
{The results are based on 1000 replications.}}
\label{tab:Example4}
\begin{adjustbox}{max width=\linewidth}
\begin{tabular}{cccccccc}
 \toprule
\multirow{2}{*}{\textbf{Method}} & 
\multirow{2}{*}{\makecell{Coverage \\ probability (\%)}} & 
\multicolumn{2}{c}{\makecell{Percentage of}} & 
\multirow{2}{*}{\makecell{Best \\ model \\ size}} & 
\multirow{2}{*}{\makecell{Selected \\ model \\ size}} & 
\multirow{2}{*}{RSS} & 
\multirow{2}{*}{MSPE} \\
\cline{3-4}
& & \makecell{false \\ negatives (\%)} & \makecell{false \\ positives (\%)} & & & & \\
  \midrule
  \multicolumn{8}{c}{ \(\theta = 2\) \quad $(n, p)$ = (100, 1640)} \\ 
GSFR &72.80 & 26.2000& 0.2705& 10.83& 9.83&  1437.56& 23.0853 \\
 PBFR &76.40 & 16.6400& 4.1423& 10.83& 68.70&  6.34& 34.2470 \\
 OGA &69.40 & 30.3800& 0.1977 & 9.59& 8.19&  1804.83& 29.9689\\
 GSFRn &76.40 & 17.7000&1.3435 & 10.83& 10.90&  264.81& 31.8009\\
 & & & & & & &  \\
 \multicolumn{8}{c}{\(\theta = 2\) \quad $(n, p)$ = (200, 3280)} \\
 GSFR &100 & 0.0000& 0.1122& 8.67& 8.67& 190.21& 1.0415 \\
 PBFR &100 & 0.0000& 0.1125& 8.67& 8.68&  189.98& 1.0463 \\
 OGA &100 & 0.0000& 0.0799 & 7.62& 7.62&  191.46& 1.0444\\
 GSFRn  &100 & 0.0000& 0.1122& 8.67& 8.67& 190.21& 1.0415 \\
 & & & & & & &  \\
  \multicolumn{8}{c}{ \(\theta = 1\) \quad $(n, p)$ = (100, 1640)} \\ 
GSFR &98.80 & 1.0200& 0.0845& 6.40& 6.36&  127.94& 2.5470 \\
 PBFR &99.10 & 0.6600& 2.9456& 6.40& 52.88&  25.98& 3.2300 \\
 OGA &97.00 & 2.9000& 0.0741 & 6.30& 6.18&  243.04& 4.0591\\
 GSFRn &99.10 & 0.6600&0.1336 & 6.40& 6.40&  91.03& 1.9622\\
 & & & & & & &  \\
 \multicolumn{8}{c}{\(\theta = 1\) \quad $(n, p)$ = (200, 3280)} \\
 GSFR &100 & 0.0000& 0.0132& 5.43& 5.43&  193.66& 1.0354 \\
 PBFR &100 & 0.0000& 0.0135& 5.43& 5.44&  193.46& 1.0384 \\
 OGA &100 & 0.0000& 0.0126 & 5.41& 5.41&  193.68& 1.0333\\
 GSFRn  &100 & 0.0000& 0.0132& 5.43& 5.43&  193.66& 1.0354 \\
  \bottomrule
\end{tabular}
\end{adjustbox}
\end{table}

Table~\ref{tab:Example4} shows that GSFR attains the lowest MSPE while maintaining a modest average model size across nearly all settings compared with OGA and PBFR, demonstrating its superior out-of-sample performance. OGA performs poorly in both in-sample and out-of-sample evaluations compared to GSFR in almost all settings. In particular, in the setting $(n,p) = (100, 1640)$ with $\theta = 2$, where within-group and inter-group correlations among predictors are high and the sample size is small, OGA attains the lowest coverage probability, resulting in the largest RSS and relatively high MSPE. PBFR occasionally attains slightly higher coverage probability than GSFR; however, consistent with the patterns observed in Example 3, it tends to overfit, selecting a large model with low in-sample RSS, while its out-of-sample MSPE always exceeds that of GSFR. Correspondingly, PBFR is consistently outperformed by GSFRn in both false positive rate and out-of-sample MSPE across all scenarios, demonstrating the greater effectiveness of GSFR's model size selection rule in eliminating redundant variables compared with BIC. As the sample size grows, GSFR's performance approaches that of GSFRn, providing empirical support for the asymptotic validity of stopping at $K_n$, which substantially reduces redundant computations. Similar asymptotic patterns in GSFR's false positive rate, MSPE, and average selected model size are observed as in Example 3, thereby corroborating Theorems~\ref{theorem1} and~\ref{theorem3}. The results from Examples 3 and 4 provide strong evidence that GSFR achieves higher efficiency and accuracy than greedy alternatives in variable selection for ultra-high dimensional data.

We next consider an example similar to that in \citet{ingStepwiseRegressionMethod2011} to evaluate the performance of GSFR relative to penalized methods. In this setting, the predictors follow a compound symmetry correlation structure, such that all variables exhibit the same degree of correlation with one another.

\noindent{\bf Example 5}.
{Consider} the following linear model:  
\begin{align}
    y_{t} = \sum_{j=1}^{q} \beta_{j} x_{t j} + \sum_{j=q+1}^{p} \beta_{j} x_{t j} + \varepsilon_{t},\quad t = 1, \ldots, n, \quad
    x_{t j} = d_{t j} + \theta \omega _{t}, \nonumber
\end{align}  
where the predictor components 
\((d_{t1}, \ldots, d_{tp}, w_t)^{\prime}\), for \(1 \le t \le n\), are independently drawn from a multivariate normal distribution with mean zero and covariance matrix \(\mathrm{I}_{p+1}\). The noise term \(\varepsilon_t\) is independently distributed as \(\mathrm{N}(0, 2.25)\) and is independent of the predictors. Within this simulation framework, we consider \((q,n,p) = (9, 200, 4000)\), with coefficient values \((\beta_1,\ldots,\beta_4) = 3.2\), \((\beta_5,\beta_6) = 4.4\), \((\beta_7,\ldots,\beta_9) = 3.5\), and set all remaining coefficients to zero. The parameter \(\theta\) is fixed at 1. Under this setting, for any \( J \subseteq\{1, \ldots, p\} \) and \( 1 \leq i \leq p \) with \( i \notin J \), we have \( \lambda_{\min }(\boldsymbol{\Gamma}(J))=\frac{1}{1+\theta^{2}}>0 \) and \( \left\|\boldsymbol{\Gamma}^{-1}(J) \mathbf{g}_{i}(J)\right\|_{1} \leq 1 \), ensuring that (\ref{assumption5_equation2}) and (\ref{assumption5_equation3}) are satisfied. Therefore, this simulation model satisfies all required assumptions.

\begin{table}[!htbp]
\centering
\caption{Variable selection results using GSFR, Lasso, Adaptive Lasso, SIS-SCAD, ISIS-SCAD and LARS methods in Example 5. 
{The results are based on 1000 replications.}}
\label{tab:Example5}
\begin{tabular}{ccccccc}
\toprule
\multirow{2}{*}{\textbf{Method}} &
\multirow{2}{*}{\makecell{Coverage \\ probability (\%)}} &
\multicolumn{2}{c}{\makecell{Percentage of}} &
\multirow{2}{*}{\makecell{Selected \\ model \\ size}} &
\multirow{2}{*}{RSS} &
\multirow{2}{*}{MSPE} \\
\cline{3-4}
& &
\makecell{false \\ negatives (\%)} &
\makecell{false \\ positives (\%)} &
& & \\
\midrule
GSFR & 100.00 & 0.0000 & 0.0152 & 9.61 & 427.51 & 2.3507 \\
Lasso & 100.00 & 0.0000 & 1.3265 & 61.94 & 324.49 & 3.3613 \\
Adaptive Lasso & 100.00 & 0.0000 & 0.0271 & 10.08 & 509.06 & 2.8799 \\
SIS-SCAD & 0.1000 & 49.0889 & 0.6654 & 31.14 & 7405.73 & 77.9723 \\
ISIS-SCAD & 100.00 & 0.0000 & 0.7016 &37.00 & 105.58 & 4.2287 \\
LARS & 100.00 & 0.0000 & 1.6963 &76.70 & 260.16 & 3.4179 \\
\bottomrule
\end{tabular}
\end{table}

In this case, GSFR successfully identifies all relevant variables in all 1000 replications and achieves the lowest false positive rate, resulting in the smallest out-of-sample MSPE while maintaining a moderate average model size. SIS-SCAD fails to include the relevant variables in most experiments, leading to the poorest in-sample RSS and out-of-sample MSPE. Although other penalized methods achieve full coverage of the relevant variables and often yield smaller in-sample RSS than GSFR, they tend to select substantially larger models and exhibit false positive rates that are significantly higher than that of GSFR. This results in inferior out-of-sample performance due to the inclusion of an excessive number of redundant variables. Among them, Adaptive Lasso yields an average model size closest to that of GSFR. Nevertheless, GSFR achieves superior in-sample and out-of-sample performance, as it effectively screens out nearly all redundant variables.  Overall, GSFR remains competitive relative to penalized alternatives while selecting substantially sparser models and achieving strong out-of-sample performance in this ultra-high dimensional setting.

\subsection{Real Data Analysis}
In this section, we illustrate the proposed methodology through two real data analysis examples. 

 \noindent{\bf Example 6 (FRED-MD Data)}. Consider first the macroeconomic variables studied by \citet{stock2002macroeconomic} and \citet{mccracken2016fred}. The original data are obtained from the FRED-MD database maintained by the Federal Reserve Bank of St.\ Louis, which is available at \href{https://research.stlouisfed.org/econ/mccracken/fred-databases/}{https://research.stlouisfed.org/econ/mccracken/fred-databases/}. As described in \citet{mccracken2016fred}, this dataset extends the widely used macroeconomic panel of \citet{stock2002macroeconomic} and covers a broad range of economic categories. These include output and income (OUT), labor market (LM), housing (HS), consumption, orders, and inventories (COI), money and credit (MC), interest and exchange rates (IER), prices (PR), and the stock market (SM). We use the preprocessed dataset from \citet{gao2025supervised}, which consists of 123 macroeconomic variables spanning the period from July 1962 to December 2019, during which all series have no missing values. The goal is to forecast the one-month-ahead Effective Federal Funds Rate (FEDFUNDS) using the remaining macroeconomic variables.
 Therefore, we have \(n = 690\) and \(p = 122\).  

In this example, we assess the performance of our method in relation to penalized approaches, including Lasso, Adaptive Lasso, SIS-SCAD, ISIS-SCAD, and LARS. The parameters are set to be the same as those used in the simulation. Before estimation, the data are standardized. Following \citet{gao2025supervised}, the sample is divided into two subsamples, with the first 80\% of observations assigned to the estimation set and the remaining 20\% reserved for out-of-sample prediction. We then employ a rolling-window scheme to produce out-of-sample forecasts. Specifically, at each step, the forecasting model is estimated using the current estimation subsample to predict the next observation of the target variable. The estimation subsample is then updated to include this new observation, and the procedure is repeated until all observations in the prediction set have been used to compute forecast errors. The average selected model size, in-sample RSS, and out-of-sample MSPE are summarized in Table \ref{tab:Example 6}.

\begin{table}[!htbp]
\centering
\caption{Variable selection results of GSFR, Lasso, Adaptive Lasso, SIS-SCAD, ISIS-SCAD, and LARS in Example 6. The results are based on rolling-window scheme.}
\label{tab:Example 6}
\begin{tabular}{cccc}
  \toprule
  \textbf{Method} & \makecell{Selected \\ model size} & RSS &  MSPE  \\
  \midrule
  GSFR & 1.00 &181.54 &0.0732    \\
  Lasso & 41.86 &121.39 &0.1246   \\
  Adaptive Lasso  & 27.20 &122.00 & 0.1543  \\
  SIS-SCAD  & 3.98 &147.38 & 0.0768  \\
  ISIS-SCAD  & 3.98 &147.38 & 0.0768  \\
  LARS  & 42.88 &120.91 & 0.1190  \\
  \bottomrule
\end{tabular}
\end{table}

Table~\ref{tab:Example 6} indicates that the model selected by GSFR attains the lowest out-of-sample MSPE while employing the fewest variables. Adaptive Lasso on average selects 27.20 variables and performs worse than the other methods in out-of-sample forecasting. LARS selects the largest model, with an average of 42.88 variables, followed by Lasso with 41.86 variables. SIS-SCAD and ISIS-SCAD select models closest in size to GSFR, with an average of 3.98 variables. The Diebold–Mariano test in \citet{diebold1995comparing} shows that GSFR's out-of-sample forecasting errors are lower than those of SIS-SCAD and ISIS-SCAD at the 5\% significance level. All of these methods select more variables than GSFR. Their in-sample errors are lower, while their out-of-sample errors are higher.
 This indicates that GSFR more accurately identifies the most significant predictors while maintaining a sparse model.

\noindent{\bf Example 7 (Tecator Infratec Food Data)}. To further evaluate the screening performance of GSFR algorithm, we consider the Meatspec data set obtained from the Tecator Infratec Food Analyzer, which is available in the R add-on package \textit{faraway} and on StatLib. The data are recorded on a Tecator Infratec Food and Feed Analyzer using near-infrared light (wavelength is 850 nm - 1050
nm) to analyze the samples. Each sample contains meat with different moisture, fat, and
protein contents. The data set consists of $n = 215$ observations. The goal is to predict the scalar fat content of a meat sample using the near-infrared absorbance measurements at 100 wavelength channels, denoted by \( \{ X_{1}, \ldots, X_{100} \} \), in the presence of spurious predictors \( \{ X_{101}, \ldots, X_{1000} \} \). Following \citet{fan2014nonparametric}, the artificial predictors are generated as follows:
\begin{equation*}
    X_{j}=\frac{Z_{j}+2U}{2}, j=101, \ldots, 1000,
\end{equation*}
where \( \{ Z_{101}, \ldots, Z_{1000} \} \) are independent standard normal random variables and \(U\) follows the standard uniform distribution.

In this example, we evaluate the performance of our methods relative to other greedy approaches. Following \citet{zhang2008twostepproceduresvariableselection}, we randomly split the data into a training set of size 50 and a test set of size 165, so that \(p/n=20\) in the training stage, suitable for fitting the ultra-high dimensional regression model~(\ref{model}). Covariates are standardized using statistics computed from the training sample only. We evaluate the average performance of GSFR, PBFR, and OGA based on 200 random experiments. The parameter \( K_n \) is chosen as \( \left\lfloor 10 \left( \frac{n}{\log p_n} \right)^{1/2} \right\rfloor \), and the adjustment terms are set to be the same as those used in the simulation. We report the average selected model size and measure in-sample performance by the residual sum of squares (RSS). Out-of-sample performance is assessed using the mean squared prediction error (MSPE), and computational efficiency is evaluated in terms of running time.

\begin{table}[!htbp]
\centering
\caption{Variable selection results of GSFR, FR, and OGA method in 
{ Example 7}. The results are based on 200 replications.}
\label{tab:Example 7}
\begin{tabular}{ccccc}
  \toprule
  \textbf{Method} & \makecell{Selected \\ model size} & RSS &  MSPE & \makecell{Running \\ time (s)} \\
  \midrule
  GSFR  & 4.54 &5.35 & 0.5867 & 0.1526 \\
   PBFR  & 20.48 &0.97 &0.6308 & 2.0471  \\
  OGA & 1.00 &35.10 &0.9376 & 0.1496   \\
  \bottomrule
\end{tabular}
\end{table}

  From the variable selection and the out-of-sample MSPE results of  the meatspec data in Table~\ref{tab:Example 7}, we see that the model selected by GSFR achieves the best MSPE on the test set while maintaining a selection time comparable to OGA. OGA selects a single variable on average, resulting in the poorest prediction performance. PBFR selects more than four times as many variables as GSFR. Although its in-sample performance is strong, the resulting out-of-sample MSPE is larger than that of GSFR. In addition, PBFR requires the longest computational time among the methods considered. GSFR requires less computational time, selects a moderate number of variables, and achieves better out-of-sample prediction performance in the ultra-high dimensional setting.

		\section{Conclusion}\label{sec5}
This paper {studied} 
a Gram–Schmidt implementation of forward regression for ultra-high dimensional linear models with highly correlated predictors. The main contribution is to show that forward-regression-type screening can be implemented in a way that is computationally attractive and numerically stable under strong multicollinearity, while retaining reliable screening performance and allowing consistent path-based model size selection. The numerical results suggest that the proposed procedure is particularly useful when sparse recovery, interpretability, and computational efficiency are all important.
 With its computational efficiency and theoretical guarantees, GSFR emerges as a reliable and effective approach for ultra-high dimensional variable selection. Although the present paper focuses on linear screening problems, the same computational idea may also be useful in other regression-based model selection settings, such as threshold estimation and structural break detection.

Current results of GSFR are established under the assumption of light-tailed distributions and 
{ the procedure uses Pearson correlation estimates, which are not robust to 
outliers. To improve the robustness of the GSFR and to handle heavy-tailed distributions, 
one may consider some 
rank correlations, such as Spearman's rho, in the selection path.}
Furthermore, 
{ instead of focusing on standard linear regression models, one 
can extend GSFR to generalized linear models (GLMs)} using the method in \citet{jiang2024decorrelated} to convert GLMs to Model (\ref{model}), thereby covering logistic, Poisson, and other models. These directions provide a basis for future work to extend and strengthen the {applicability} of GSFR.


\section*{Supplementary Material}
This supplementary material includes theoretical proofs of the main theorems.
		
		
\section*{Disclosure Statement}
We declare that there are no relevant financial or non-financial competing interests to disclose.

        \section*{Data Availability Statement}
        The authors confirm that the first dataset supporting the findings of this study is from the FRED-MD database maintained by the St.\ Louis Fed at \href{https://research.stlouisfed.org/econ/mccracken/fred-databases/}{https://research.stlouisfed.org/econ/mccracken/fred-databases/}, while the second dataset can be obtained from the R package \texttt{faraway} and StatLib.
		

%
	
		\singlespacing
\bibliographystyle{econometrica-3}
\let\oldbibliography\thebibliography
\renewcommand{\thebibliography}[1]{%
  \oldbibliography{#1}%
  \setlength{\itemsep}{3pt}%
}
	{\footnotesize
		\bibliography{GSFR-BIB}

@article{diebold1995comparing,
  title={Comparing Predictive Accuracy},
  author={Diebold, Francis X and Mariano, Roberto S},
  journal={Journal of Business \& Economic Statistics},
  volume={13},
  number={3},
  pages={253--263},
  year={1995},
  publisher={Taylor \& Francis}
}

@article{barron2008approximation,
  title={Approximation and learning by greedy algorithms},
  author={Barron, Andrew R and Cohen, Albert and Dahmen, Wolfgang and DeVore, Ronald A},
   journal = {The Annals of Statistics},
  volume = {36},
  number = {},
  pages = {64--94},
  year={2008}
}

@inproceedings{donoho2006breakdown,
  title={Breakdown point of model selection when the number of variables exceeds the number of observations},
  author={Donoho, David and Stodden, Victoria},
  booktitle={The 2006 IEEE international joint conference on neural network proceedings},
  pages={1916--1921},
  year={2006},
  organization={IEEE}
}

@article{borodinProjectionGreedyAlgorithm2021,
  title = {Projection Greedy Algorithm},
  author = {Borodin, P. A. and Konyagin, S. V.},
  year = {2021},
  month = jul,
  journal = {Mathematical Notes},
  volume = {110},
  number = {1},
  pages = {16--25},
  issn = {1573-8876},
  doi = {10.1134/S0001434621070026}
}

@article{ingStepwiseRegressionMethod2011,
  title = {A Stepwise Regression Method and Consistent Model Selection for High-Dimensional Sparse Linear Models},
  author = {Ing, Ching-Kang and Lai, Tze Leung},
  year = {2011},
  month = oct,
  journal = {Statistica Sinica},
  volume = {21},
  number = {4},
  issn = {10170405},
  doi = {10.5705/ss.2010.081},
  urldate = {2024-10-21},
  langid = {english}
}

@article{temlyakov2000weak,
  title={Weak greedy algorithms},
  author={Temlyakov, Vladimir N},
  journal={Advances in Computational Mathematics},
  volume={12},
  number={2},
  pages={213--227},
  year={2000},
  publisher={Springer}
}

@article{ledoit2004well,
  title={A well-conditioned estimator for large-dimensional covariance matrices},
  author={Ledoit, Olivier and Wolf, Michael},
  journal={Journal of multivariate analysis},
  volume={88},
  number={2},
  pages={365--411},
  year={2004},
  publisher={Elsevier}
}

@article{10.1214/009053607000000758,
author = {Peter J. Bickel and Elizaveta Levina},
title = {{Regularized estimation of large covariance matrices}},
volume = {36},
journal = {The Annals of Statistics},
number = {1},
publisher = {Institute of Mathematical Statistics},
pages = {199 -- 227},
year = {2008},
doi = {10.1214/009053607000000758},
URL = {https://doi.org/10.1214/009053607000000758}
}

@article{tibshirani1996regression,
  title={Regression shrinkage and selection via the lasso},
  author={Tibshirani, Robert},
  journal={Journal of the Royal Statistical Society Series B: Statistical Methodology},
  volume={58},
  number={1},
  pages={267--288},
  year={1996},
  publisher={Oxford University Press}
}

@article{fan2001variable,
  title={Variable selection via nonconcave penalized likelihood and its oracle properties},
  author={Fan, Jianqing and Li, Runze},
  journal={Journal of the American statistical Association},
  volume={96},
  number={456},
  pages={1348--1360},
  year={2001},
  publisher={Taylor \& Francis}
}

@article{zou2005regularization,
  title={Regularization and variable selection via the elastic net},
  author={Zou, Hui and Hastie, Trevor},
  journal={Journal of the Royal Statistical Society Series B: Statistical Methodology},
  volume={67},
  number={2},
  pages={301--320},
  year={2005},
  publisher={Oxford University Press}
}

@article{Peter2006,
author = {Peter B{\"u}hlmann},
title = {{Boosting for high-dimensional linear models}},
volume = {34},
journal = {The Annals of Statistics},
number = {2},
publisher = {Institute of Mathematical Statistics},
pages = {559 -- 583},
year = {2006},
doi = {10.1214/009053606000000092},
URL = {https://doi.org/10.1214/009053606000000092}
}

@article{wang2009forward,
  title={Forward regression for ultra-high dimensional variable screening},
  author={Wang, Hansheng},
  journal={Journal of the American Statistical Association},
  volume={104},
  number={488},
  pages={1512--1524},
  year={2009},
  publisher={Taylor \& Francis}
}

@article{fan2008sure,
  title={Sure independence screening for ultrahigh dimensional feature space},
  author={Fan, Jianqing and Lv, Jinchi},
  journal={Journal of the Royal Statistical Society Series B: Statistical Methodology},
  volume={70},
  number={5},
  pages={849--911},
  year={2008},
  publisher={Oxford University Press}
}

@article{ing2020model,
  title={Model selection for high-dimensional linear regression with dependent observations},
  author={Ing, Ching-Kang},
  journal={The Annals of Statistics},
  volume={48},
  number={4},
  pages={1959--1980},
  year={2020},
  publisher={JSTOR}
}

@article{huang2008asymptotic,
  title={Asymptotic properties of bridge estimators in sparse high-dimensional regression models},
  author={Huang, J. and Horowitz, J. L. and Ma, S.},
  journal={The Annals of Statistics},
  volume={36},
  number={2},
  pages={587--613},
  year={2008},
}

@article{jiang2024decorrelated,
  title={Decorrelated forward regression for high dimensional data analysis},
  author={Jiang, Xuejun and Ma, Yue and Wang, Haofeng},
  journal={arXiv preprint arXiv:2408.12272},
  year={2024}
}

@misc{zhang2008twostepproceduresvariableselection,
      title={Some Two-Step Procedures for Variable Selection in High-Dimensional Linear Regression}, 
      author={Jian Zhang and Xinge Jessie Jeng and Han Liu},
      year={2008},
      eprint={0810.1644},
      archivePrefix={arXiv},
      primaryClass={math.ST},
      url={https://arxiv.org/abs/0810.1644}, 
}

@article{fan2014nonparametric,
  title={Nonparametric independence screening in sparse ultra-high-dimensional varying coefficient models},
  author={Fan, Jianqing and Ma, Yunbei and Dai, Wei},
  journal={Journal of the American Statistical Association},
  volume={109},
  number={507},
  pages={1270--1284},
  year={2014},
  publisher={Taylor \& Francis}
}

@article{JMLR:v25:23-0882,
  author  = {Shuo-Chieh Huang and Ruey S. Tsay},
  title   = {Scalable High-Dimensional Multivariate Linear Regression for Feature-Distributed Data},
  journal = {Journal of Machine Learning Research},
  year    = {2024},
  volume  = {25},
  number  = {205},
  pages   = {1--59},
  url     = {http://jmlr.org/papers/v25/23-0882.html}
}

@article{stock2002macroeconomic,
  title={Macroeconomic forecasting using diffusion indexes},
  author={Stock, James H and Watson, Mark W},
  journal={Journal of business \& economic statistics},
  volume={20},
  number={2},
  pages={147--162},
  year={2002},
  publisher={Taylor \& Francis}
}

@article{mccracken2016fred,
  title={FRED-MD: A monthly database for macroeconomic research},
  author={McCracken, Michael W and Ng, Serena},
  journal={Journal of Business \& Economic Statistics},
  volume={34},
  number={4},
  pages={574--589},
  year={2016},
  publisher={Taylor \& Francis}
}

@article{gao2025supervised,
  title={Supervised dynamic pca: Linear dynamic forecasting with many predictors},
  author={Gao, Zhaoxing and Tsay, Ruey S},
  journal={Journal of the American Statistical Association},
  volume={120},
  number={550},
  pages={869--883},
  year={2025},
  publisher={Taylor \& Francis}
}
		}
		
\end{document}





\def\spacingset#1{\renewcommand{\baselinestretch}%
{#1}\small\normalsize} \spacingset{1}


{
 
\title{Supplementary Material for ``Forward Regression via Gram–Schmidt Orthogonalization for Ultra-High Dimensional Linear Models"}
\author{
}

 \date{}

\maketitle
}


\spacingset{1.75} 


\abovedisplayskip=0.1pt
\belowdisplayskip=0.1pt


\setcounter{section}{0}
	\setcounter{subsection}{0}

 	\renewcommand{\thesection}{\Alph{section}}

	\renewcommand{\thesubsection}{\thesection.\arabic{subsection}}
	
	\setcounter{equation}{0}
	\renewcommand{\theequation}{\thesection.\arabic{equation}}
	

	
	\renewcommand{\theequation}{S.\arabic{equation}}%
	\renewcommand{\thefigure}{S.\arabic{figure}} \setcounter{figure}{0}
	\renewcommand{\thetable}{S.\Roman{table}} \setcounter{table}{0}
	
	

	
	

This supplementary material includes theoretical proofs of the main theorems. 


\section{Proof of Theorem \ref{theorem1} }
Similar to \citet{ingStepwiseRegressionMethod2011}, we replace \( x_{tj} \) with \( x_{tj} / \sigma_j \) and \( x_j \) with \( x_j / \sigma_j \) in both GSFR and its population version. This standardization allows us to assume, without loss of generality, that \( \sigma_j = 1 \), for all \( 1 \leq j \leq p_n \), leading to \( z_j = x_j \).

\subsection{Supporting lemmas }
The following lemmas, adapted from \citet{ingStepwiseRegressionMethod2011} to fit the GSFR framework, play a key role in proving Theorem \ref{theorem1}. Detailed proofs can be found in the supplementary material of \citet{ingStepwiseRegressionMethod2011}.

\begin{lemma}
With the same notation and assumptions as {those} in Theorem \ref{theorem1}, 
there exists $C>0$ such that
\begin{equation}
\max _{1 \leq i, j \leq p_{n}} P\left\{\left|\sum_{t=1}^{n}\left(x_{t i} x_{t j}-\sigma_{i j}\right)\right|>n \delta_{n}\right\} \leq \exp \left(-C n \delta_{n}^{2}\right)
\nonumber
\end{equation}
for all large $n$, where $\sigma_{i j}=\operatorname{Cov}\left(x_{i}, x_{j}\right)$ and $\delta_{n}$ are positive constants satisfying $\delta_{n} \rightarrow 0$ and $n \delta_{n}^{2} \rightarrow \infty$ as $n \rightarrow \infty$. Define \(\boldsymbol{\Gamma}(J)\) as in  (\ref{assumption5_equation1}), and let \(\hat{\boldsymbol{\Gamma}}_{n}(J)\) be the corresponding sample covariance matrix.
Then, for all large $n$,
\begin{equation}
P\left\{\max _{1 \leq \sharp(J) \leq K_{n}}\left\|\hat{\boldsymbol{\Gamma}}_{n}(J)-\boldsymbol{\Gamma}(J)\right\|>K_{n} \delta_{n}\right\} \leq p_{n}^{2} \exp \left(-C n \delta_{n}^{2}\right).
\nonumber
\end{equation}
If furthermore $K_{n} \delta_{n}=O(1)$, then there exists $c>0$ such that
\begin{equation}
P\left\{\max _{1 \leq \sharp(J) \leq K_{n}}\left\|\hat{\boldsymbol{\Gamma}}_{n}^{-1}(J)-\boldsymbol{\Gamma}^{-1}(J)\right\|>K_{n} \delta_{n}\right\} \leq p_{n}^{2} \exp \left(-c n \delta_{n}^{2}\right)
\nonumber
\end{equation}
for all large $n$, where $\hat{\boldsymbol{\Gamma}}_{n}^{-1}$ denotes a generalized inverse when $\hat{\boldsymbol{\Gamma}}_{n}$ is singular.
\label{lemma1}
\end{lemma}
\par

\begin{lemma}
With the same notation and assumptions as 
{those} in Lemma \ref{lemma1}, let $\delta_{n}$ be positive numbers such that $\delta_{n}=O\left(n^{-\theta}\right)$ for some $0<\theta<1 / 2$ and $n \delta_{n}^{2} \rightarrow \infty$. Then there exists $\eta>0$ such that for all large $n$,
\begin{equation}
\max _{1 \leq i \leq p_{n}} P\left(\left|\sum_{t=1}^{n} \varepsilon_{t} x_{t i}\right| \geq n \delta_{n}, \Omega_{n}\right) \leq \exp \left(-\eta n \delta_{n}^{2}\right),
\nonumber
\end{equation}
where $\Omega_{n}=\left\{\max _{1 \leq t \leq n}\left|\varepsilon_{t}\right|<(\log n)^{2}\right\}$. It follows from assumption (\ref{assumption2}) that $\lim _{n \rightarrow \infty} P\left(\Omega_{n}^{c}\right)=0$.
\label{lemma2}
\end{lemma}
\par

\begin{lemma}
With the same notation and assumptions as {those} in Lemma \ref{lemma1}, and using Lemmas \ref{lemma1} and \ref{lemma2}, the following result holds for some $\lambda>0$,
\begin{equation}
\begin{array}{l}
P\left(\max _{\sharp(J) \leq K_{n}-1, i \notin J}\left|n^{-1} \sum_{t=1}^{n} \varepsilon_{t} \hat{x}_{t i ; J}^{\perp}\right|>\lambda\left(\frac{\log p_{n} }{n}\right)^{1 / 2}, \Omega_{n}\right)=o(1), \\
P\left(\max _{i, j \notin J, \sharp(J) \leq K_{n}-1}\left|n^{-1} \sum_{t=1}^{n} x_{t j} \hat{x}_{t i ; J}^{\perp}-E\left(x_{j} x_{i ; J}^{\perp}\right)\right|>\lambda\left(\frac{\log p_{n} }{n}\right)^{1 / 2}\right)=o(1).
\end{array}
\nonumber
\end{equation}
\label{lemma3}
\end{lemma}
\par

\begin{lemma}
We cite Theorem 2 from \citet{borodinProjectionGreedyAlgorithm2021} to demonstrate the convergence rate of the semi-population version of GSFR. Let
\begin{equation*}
 x_{0} \in \mathcal{D}^{\lambda}:=\overline{\left\{\sum_{k=1}^{m} \lambda_{k} s_{k}: s_{k} \in D, m \in \mathbb{N}, \sum_{k=1}^{m}\left|\lambda_{k}\right| \leq \lambda\right\}}.
\end{equation*}
Then
\begin{equation}
\left|x_{n}\right| \leq \frac{\lambda}{\left(1+\sum_{j=0}^{n-1} t_{j}^{2}\right)^{1 / 2}}.
\nonumber
\end{equation}
\label{lemma4}
\end{lemma}

\subsection{Proof}

\begin{proof}
To avoid the possible numerical problem caused by dividing extremely small $\left(\sum_{t=1}^{n} \left(\hat{x}^{\bot}_{t i; J}\right)^{2}\right)^{1/2}$, we prove a more general result  using 
\begin{equation*}
	\hat{\mu}_{J, i} = \frac{n^{-1} \sum_{t=1}^{n} \left(y_{t} - \hat{y}_{t; J}\right) \hat{x}_{t i; J}^{\bot}}{\left(n^{-1} \sum_{t=1}^{n} \left(\hat{x}^{\bot}_{t i; J}\right)^{2}\right)^{1/2}+\rho_{1}\left ( \frac{\log p_{n} }{n}  \right ) ^{1/2} },
\end{equation*}
where constant $\rho_{1} \ge 0$.
For $i \notin J$, we further simplify (\ref{ref3}):

\begin{equation}
\mu_{J, i}=\frac{\sum_{j \notin J} \beta_{j} E\left[\left(x_{j}-x_{j}^{(J)}\right) x_{i}\right]}{E\left[{x_{i;J}^{\bot^{2}}}\right]^{1/2}} =\frac{\sum_{j \notin J} \beta_{j} E\left[x_{j}\left(x_{i}-x_{i}^{(J)}\right)\right]}{E\left[{x_{i;J}^{\bot^{2}}}\right]^{1/2}} =\frac{\sum_{j \notin J} \beta_{j} E\left(x_{j} x_{i ; J}^{\perp}\right)}{E\left[{x_{i;J}^{\bot^{2}}}\right]^{1/2}}.
\label{ref6}
\end{equation}

Since $y_{t}=\sum_{j=1}^{p_{n}} \beta_{j} x_{t j}+\varepsilon_{t}$ and $\hat{y}_{t;J} = \sum_{i\in J}\beta _{i}x_{ti} + \sum_{i\notin J}\beta _{i}\hat{x}_{ti;J}+\hat{\varepsilon }_{t;J}$, where $\hat{\varepsilon }_{t;J}$ and $\hat{x}_{t i ; J}$ represent the fitted values of $\varepsilon_{t}$ and $x_{t i}$, respectively, and since 
\(
\sum_{t=1}^{n}\left(\varepsilon_{t}-\hat{\varepsilon}_{t ; J}\right) x_{t i}=\sum_{t=1}^{n} \varepsilon_{t} \hat{x}_{t i ; J}^{\perp}
\),
by applying (\ref{ref3}) and (\ref{ref6}), it follows that 

\begin{equation}
\begin{aligned}
    \hat{\mu}_{J, i} - \mu_{J, i} 
    &= \frac{n^{-1} \sum_{t=1}^{n} \varepsilon_{t} \hat{x}_{t i; J}^{\perp}}
    {\left(n^{-1} \sum_{t=1}^{n} \hat{x}_{t i; J}^{\bot^2}\right)^{1/2}
    +\rho_{1}\left ( \frac{\log p_{n} }{n}  \right ) ^{1/2} }  \\
    &\quad + \sum_{j \notin J} \beta_{j} \left\{ 
    \frac{n^{-1} \sum_{t=1}^{n} x_{t j} \hat{x}_{t i; J}^{\perp}}
    {\left(n^{-1} \sum_{t=1}^{n} \hat{x}_{t i; J}^{\bot^2}\right)^{1/2}
    +\rho_{1}\left ( \frac{\log p_{n} }{n}  \right ) ^{1/2} } 
    - \frac{E\left(x_{j} x_{i; J}^{\perp}\right)}
    {E\left(x_{i; J}^{\perp^2}\right)^{1/2}} \right\}.
\end{aligned}
\label{ref7}
\end{equation}

We now derive probability bounds for the right-hand side of (\ref{ref7}).  
First, we establish a result similar to (A.15) in \cite{ingStepwiseRegressionMethod2011}.  
By Lemma \ref{lemma3} and for the same $\lambda > 0$,  

\begin{align}
    &P\Bigg(\max_{i \notin J, \sharp(J) \leq K_{n}-1}  
    \Big| n^{-1} \sum_{t=1}^{n} \hat{x}_{t i; J}^{\perp^2}  
    - E\left(x_{i; J}^{\perp^2}\right) \Big|  
    > \lambda \left ( \frac{\log p_{n} }{n}  \right ) ^{1/2} \Bigg) \nonumber \\
    &= P\Bigg(\max_{i \notin J, \sharp(J) \leq K_{n}-1}  
    \Big| n^{-1} \sum_{t=1}^{n} x_{t i} \hat{x}_{t i; J}^{\perp}  
    - E\left(x_{i} x_{i; J}^{\perp}\right) \Big|  
    > \lambda \left ( \frac{\log p_{n} }{n}  \right ) ^{1/2} \Bigg) \nonumber \\
    &\le P\Bigg(\max_{i, j \notin J, \sharp(J) \leq K_{n}-1}  
    \Big| n^{-1} \sum_{t=1}^{n} x_{t j} \hat{x}_{t i; J}^{\perp}  
    - E\left(x_{j} x_{i; J}^{\perp}\right) \Big|  
    > \lambda \left ( \frac{\log p_{n} }{n}  \right ) ^{1/2} \Bigg) = o(1).
    \label{collary1}
\end{align}

Since (\ref{assumption5_equation2}) assumes $\min_{1 \leq \sharp(J) \leq K_{n}} \lambda_{\min }(\boldsymbol{\Gamma}(J))>\delta$ for some small $\delta>0$, it follows that $x_{t1} , \dots, x_{tp_{n}}$ are linearly independent and so are $x_{1} , \dots, x_{p_{n}}$. 

Note for $i \notin J$,

\begin{equation*}
	x_{i: J}^{\perp}=x_{i}-x_{i}^{(J)}=x_{i}-\beta^{(i;J)}_{1}x_{1^{(J)}}-\beta^{(i;J)}_{2}x_{2^{(J)}}-\dots- \beta^{(i;J)}_{\sharp(J)}x_{\sharp(J)^{(J)}},
\end{equation*}

here $\{x_{k^{(J)}}\}_{k=1}^{\sharp(J)}$ are the variables with indices in $J$ (ordered increasingly), and $\beta^{(i;J)}_{k}$ are their corresponding projection coefficients.

Then

\begin{align}
    E\left ( x_{i: J}^{\perp^{2}} \right ) &= E\left (\left ( x_{i}-x_{i}^{(J)} \right )^{2} \right ) \nonumber \\
    &= E\Bigg( \Big( x_{i} - \beta^{(i;J)}_{1} x_{1^{(J)}}  
    - \beta^{(i;J)}_{2} x_{2^{(J)}} - \dots  
    - \beta^{(i;J)}_{\sharp(J)} x_{\sharp(J)^{(J)}} \Big)^{2} \Bigg).
    \label{ref8}
\end{align}

Define $(\sharp(J)+1) \times 1$ column  

\begin{equation*}
	\mathbf{\beta}^{(i;J)} = \begin{bmatrix}
1 \quad (\text{row } 1) \\
-\beta^{(i;J)}_{k} \quad (\text{row } k+1, k=1,\ldots,\sharp(J)) \\
\end{bmatrix},
\end{equation*}
then the equation (\ref{ref8}) can be represented in the quadratic form
\[
\mathbf{\beta}^{(i;J)\top} \boldsymbol{\Gamma}(J\cup\{i\}) \mathbf{\beta}^{(i;J)},
\]
where, without loss of generality, the variable ordering in the construction of $\boldsymbol{\Gamma}(J\cup\{i\})$ is consistent with that in the coefficient vector $\mathbf{\beta}^{(i;J)}$. We have
\begin{equation}
    \mathbf{\beta}^{(i;J)^{\prime }}\boldsymbol{\Gamma}(J\cup\{i\} )\mathbf{\beta}^{(i;J)} 
    \ge \frac{\mathbf{\beta}^{(i;J)^{\prime }}\boldsymbol{\Gamma}(J\cup\{i\} )\mathbf{\beta}^{(i;J)}}{\mathbf{\beta}^{(i;J)^{\prime }}\mathbf{\beta}^{(i;J)}}  
    \ge \min_{\mathbf{x}\ne \mathbf{0} } \frac{\mathbf{x^{\prime }}\boldsymbol{\Gamma}(J\cup\{i\} )\mathbf{x}}{\mathbf{x}^{\prime }\mathbf{x}} 
    =\lambda_{\min }(\boldsymbol{\Gamma}(J\cup\{i\} ) )> \delta >0,
    \label{collary2}
    \nonumber
\end{equation}
which proves $E\left ( x_{i: J}^{\perp^{2}} \right) > \delta >0$ for $i \notin J $ and $\sharp(J) \leq K_{n}-1$. Note by assumption 

\begin{equation*}
	E\left ( x_{i: J}^{\perp^{2}} \right) = E\left ( x_{i}^{2}\right) - E\left ( x_{i}^{(J)^{2}}\right)\le E\left ( x_{i}^{2}\right) = 1 ,
\end{equation*}
we summarize that for $i \notin J $

\begin{equation}
	\delta < E\left ( x_{i: J}^{\perp^{2}} \right) \le 1.
	\label{collary2}
\end{equation}

Combing (\ref{collary1}) and (\ref{collary2}) we now show that there exists a positive constant $s>0$, independent of m and n, such that

\begin{equation}
\begin{aligned}
\lim_{n \rightarrow \infty} P\left(A_{n}^{c}\left(K_{n}\right)\right) &= 0, \text{ where} \\
A_{n}(m) &= \left\{\max_{(J, i): \sharp(J) \leq m-1, i \notin J}\left|\hat{\mu}_{J, i}-\mu_{J, i}\right| \leq s\left(\frac{\log p_{n}}{n}\right)^{1/2}\right\}.
\end{aligned}
\label{ref9}
\end{equation}\\

To prove (\ref{ref9}), it suffices to show that for some $\lambda^{*}>0$,
\begin{equation}
	P\left(\max _{\sharp(J) \leq K_{n}-1, i \notin J}\left|\frac{n^{-1} \sum_{t=1}^{n} \varepsilon_{t} \hat{x}_{t i; J}^{\perp}}{\left(n^{-1} \sum_{t=1}^{n} \hat{x}_{t i; J}^{\bot^2}\right)^{1/2}+\rho_{1}\left ( \frac{\log p_{n} }{n}  \right ) ^{1/2}}\right|>\lambda^{*}\left ( \frac{\log p_{n} }{n}  \right ) ^{1/2}, \Omega_{n}, \Omega^{\prime}_{n}\right)=o(1),
	\label{ref10}
\end{equation}
\\
\begin{equation}
\begin{split}
    P\Bigg(&\max_{\sharp(J) \leq K_{n}-1, i,j \notin J} 
    \Bigg| \sum_{j \notin J} \beta_{j} \Bigg\{ \frac{n^{-1} \sum_{t=1}^{n} x_{t j} \hat{x}_{t i; J}^{\perp}} 
    {\left(n^{-1} \sum_{t=1}^{n} \hat{x}_{t i; J}^{\bot^2}\right)^{1/2} + \rho_{1} \left( \frac{\log p_{n}}{n} \right)^{1/2}} \\
    &\quad\quad - \frac{E\left(x_{j} x_{i; J}^{\perp}\right)}{E\left(x_{i; J}^{\perp^2}\right)^{1/2}} \Bigg\} \Bigg| 
    > \lambda^{*} \left( \frac{\log p_{n}}{n} \right)^{1/2}, \Omega^{\prime}_{n} \Bigg) = o(1),
\end{split}
\label{ref11}
\end{equation}

where 
\[\Omega^{\prime}_{n}=\left \{ \max_{i \notin J, \sharp(J) \leq K_{n}-1} \left| n^{-1} \sum_{t=1}^{n} \hat{x}_{t i; J}^{\perp^2} - E\left(x_{i; J}^{\perp^2}\right)\right| \le \lambda \left ( \frac{\log p_{n} }{n}  \right ) ^{1/2} \right \}.\]

We now prove

\begin{equation}
	P\left(\max _{\sharp(J) \leq K_{n}-1, i \notin J}\left|\frac{n^{-1} \sum_{t=1}^{n} \varepsilon_{t} \hat{x}_{t i; J}^{\perp}}{\left(n^{-1} \sum_{t=1}^{n} \hat{x}_{t i; J}^{\bot^2}\right)^{1/2}+\rho_{1}\left ( \frac{\log p_{n} }{n}  \right ) ^{1/2}}\right|>\lambda_{1}\left ( \frac{\log p_{n} }{n}  \right ) ^{1/2}, \Omega_{n},  \Omega^{\prime}_{n}\right)=o(1),
	\label{ref12}
    \nonumber
\end{equation}
where \( \delta^{\prime} = \frac{\delta}{4} \) and \( \lambda _{1} =\frac{\lambda }{{\delta^{\prime}}^{\frac{1}{2}}} \).

Since for sufficiently large n we have $ n^{-1} \sum_{t=1}^{n} \hat{x}_{t i; J}^{\perp^2} > \delta^{\prime}>0$ on $\Omega^{\prime}_{n}$, it suffices to prove

\begin{equation}
	P\left(\max _{\sharp(J) \leq K_{n}-1, i \notin J}\left|n^{-1} \sum_{t=1}^{n} \varepsilon_{t} \hat{x}_{t i; J}^{\perp}\right|>\lambda_{1}{\delta^{\prime}}^{\frac{1}{2}}\left ( \frac{\log p_{n} }{n}  \right ) ^{1/2}, \Omega_{n},  \Omega^{\prime}_{n}\right)=o(1).
    \nonumber
\end{equation} 

Using Lemma \ref{lemma3}, we have
\begin{align}
    &P\left(\max _{\sharp(J) \leq K_{n}-1, i \notin J}\left|n^{-1} \sum_{t=1}^{n} \varepsilon_{t} \hat{x}_{t i; J}^{\perp}\right|>\lambda_{1}{\delta^{\prime}}^{\frac{1}{2}}\left ( \frac{\log p_{n} }{n}  \right ) ^{1/2}, \Omega_{n},  \Omega^{\prime}_{n}\right) \nonumber \\
    &= P\left(\max _{\sharp(J) \leq K_{n}-1, i \notin J}\left|n^{-1} \sum_{t=1}^{n} \varepsilon_{t} \hat{x}_{t i; J}^{\perp}\right|>\lambda\left ( \frac{\log p_{n} }{n}  \right ) ^{1/2}, \Omega_{n},  \Omega^{\prime}_{n}\right) \nonumber \\
    &\le P\left(\max _{\sharp(J) \leq K_{n}-1, i \notin J}\left|n^{-1} \sum_{t=1}^{n} \varepsilon_{t} \hat{x}_{t i; J}^{\perp}\right|>\lambda\left ( \frac{\log p_{n} }{n}  \right ) ^{1/2}, \Omega_{n} \right)
        \nonumber \\
    & = o(1).
    \nonumber
\end{align}

Using (\ref{assumption4}) we let $M^{\prime}=\sup _{n \geq 1} \sum_{j=1}^{p_{n}}\left|\beta_{j} \sigma_{j}\right|$ and $\lambda _{2} = \frac{2\lambda M^{\prime }}{\delta^{\prime^{\frac{3}{2}}}}+\frac{\rho_{1} M^{\prime } }{\delta^{\prime}} $, we now prove

\begin{equation}
\begin{split}
    P\Bigg(&\max_{\sharp(J) \leq K_{n}-1, i,j \notin J} 
    \Bigg| \sum_{j \notin J} \beta_{j} \left\{
    \frac{n^{-1} \sum_{t=1}^{n} x_{t j} \hat{x}_{t i; J}^{\perp}} 
    {\left(n^{-1} \sum_{t=1}^{n} \hat{x}_{t i; J}^{\bot^2}\right)^{1/2} + \rho_{1} \left( \frac{\log p_{n}}{n} \right)^{1/2}} \right. \\
    & \quad - \left. \frac{E\left(x_{j} x_{i; J}^{\perp}\right)}{E\left(x_{i; J}^{\perp^2}\right)^{1/2}} \right\} \Bigg| > \lambda_{2} \left( \frac{\log p_{n}}{n} \right)^{1/2}, \Omega^{\prime}_{n} \Bigg) = o(1).
\end{split}
\label{ref13}
\end{equation}

We first deal with the denominators:

\begin{align*}
    P\Bigg(&\max _{\sharp(J) \leq K_{n}-1, i,j \notin J} 
    \Bigg|\sum_{j \notin J} \beta_{j} 
    \Bigg\{ \frac{n^{-1} \sum_{t=1}^{n} x_{t j} \hat{x}_{t i; J}^{\perp}} 
    {\left(n^{-1} \sum_{t=1}^{n} \hat{x}_{t i; J}^{\bot^2}\right)^{1/2} 
    + \rho_{1} \left( \frac{\log p_{n} }{n} \right)^{1/2}} \\
    &\quad\quad - \frac{E\left(x_{j} x_{i; J}^{\perp}\right)} 
    {E\left(x_{i; J}^{\perp^2}\right)^{1/2}} 
    \Bigg\} \Bigg| > \lambda_{2} \left( \frac{\log p_{n} }{n} \right)^{1/2}, 
    \Omega^{\prime}_{n} \Bigg) \\
    &\leq P\Bigg(\max _{\sharp(J) \leq K_{n}-1, i,j \notin J} 
    \Bigg| \frac{n^{-1} \sum_{t=1}^{n} x_{t j} \hat{x}_{t i; J}^{\perp}} 
    {\left(n^{-1} \sum_{t=1}^{n} \hat{x}_{t i; J}^{\bot^2}\right)^{1/2} 
    + \rho_{1} \left( \frac{\log p_{n} }{n} \right)^{1/2}} \\
    &\quad\quad - \frac{E\left(x_{j} x_{i; J}^{\perp}\right)} 
    {E\left(x_{i; J}^{\perp^2}\right)^{1/2}} 
    \Bigg| > \left( \frac{2\lambda }{\delta^{\prime^{3/2}}} 
    + \frac{\rho_{1} }{\delta^{\prime}}  \right) 
    \left( \frac{\log p_{n} }{n} \right)^{1/2}, 
    \Omega^{\prime}_{n} \Bigg) \\
    &\leq P\Bigg(\max _{\sharp(J) \leq K_{n}-1, i,j \notin J} 
    E\left(x_{i; J}^{\perp^2}\right)^{1/2} 
    \Bigg| n^{-1} \sum_{t=1}^{n} x_{t j} \hat{x}_{t i; J}^{\perp} 
    - E\left(x_{j} x_{i; J}^{\perp}\right) \Bigg| \\
    &\quad + \max _{\sharp(J) \leq K_{n}-1, i,j \notin J}  
    \Bigg| \left(n^{-1} \sum_{t=1}^{n} \hat{x}_{t i; J}^{\bot^2}\right)^{1/2} 
    + \rho_{1} \left( \frac{\log p_{n} }{n} \right)^{1/2} 
    - E\left(x_{i; J}^{\perp^2}\right)^{1/2} \Bigg| \\
    &\quad\quad \times \Bigg| E\left(x_{j} x_{i; J}^{\perp}\right) \Bigg| 
    > \left( \frac{2\lambda }{\delta^{\prime^{1/2}}} + \rho_{1} \right) 
    \left( \frac{\log p_{n} }{n} \right)^{1/2}, \Omega^{\prime}_{n} \Bigg).
\end{align*}

Since $\left | E\left(x_{j} x_{i; J}^{\perp}\right) \right | = \left| E\left(x_{j; J}^{\perp}x_{i; J}^{\perp}\right) \right |\le 1$ and $ \max_{i \notin J, \sharp(J) \leq K_{n}-1} \left| n^{-1} \sum_{t=1}^{n} \hat{x}_{t i; J}^{\perp^2} - E\left(x_{i; J}^{\perp^2}\right)\right| \le \lambda \left ( \frac{\log p_{n} }{n}  \right ) ^{1/2}$ on $\Omega^{\prime}_{n}$, by Lemma \ref{lemma3} we have

\begin{align}
    \text{RHS} &\le P\Bigg(\max_{\sharp(J) \leq K_{n}-1, i,j \notin J} 
    \Bigg| n^{-1} \sum_{t=1}^{n} x_{t j} \hat{x}_{t i; J}^{\perp} 
    - E\left(x_{j} x_{i; J}^{\perp}\right) \Bigg| \nonumber \\
    &\quad\quad + \frac{\lambda}{ \delta^{\prime^{1/2}}} 
    \left( \frac{\log p_{n}}{n} \right)^{1/2} 
    > \frac{2\lambda}{\delta^{\prime^{1/2}}} \left( \frac{\log p_{n}}{n} \right)^{1/2}, \Omega^{\prime}_{n} \Bigg) \nonumber \\
    &= P\left(\max _{\sharp(J) \leq K_{n}-1, i,j \notin J} \left|n^{-1} \sum_{t=1}^{n} x_{t j} \hat{x}_{t i; J}^{\perp}-E\left(x_{j} x_{i; J}^{\perp}\right) \right|
         > \frac{\lambda}{\delta^{\frac{1}{2}}}\left ( \frac{\log p_{n} }{n}  \right ) ^{1/2}, \Omega^{\prime}_{n} \right)\nonumber \\
    &\le P\left(\max _{\sharp(J) \leq K_{n}-1, i,j \notin J} \left|n^{-1} \sum_{t=1}^{n} x_{t j} \hat{x}_{t i; J}^{\perp}-E\left(x_{j} x_{i; J}^{\perp}\right) \right|
         >\lambda \left ( \frac{\log p_{n} }{n}  \right ) ^{1/2}\right)\nonumber \\
    &=o(1)\nonumber
\end{align}

Let $\lambda^{*} = \max\left \{ \lambda_{1} ,   \lambda_{2} \right \} $ , then (\ref{ref10}) and (\ref{ref11}) hold using $\lambda^{*}$. Let $s=2\lambda^{*}$,

\begin{align}
	&P\left(\max_{(J, i): \sharp(J) \leq m-1, i \notin J}\left|\hat{\mu}_{J, i}-\mu_{J, i}\right| > s\left(\frac{\log p_{n}}{n}\right)^{1/2}\right)
	\nonumber \\
	&\le P\left(\max _{\sharp(J) \leq K_{n}-1, i \notin J}\left|\frac{n^{-1} \sum_{t=1}^{n} \varepsilon_{t} \hat{x}_{t i; J}^{\perp}}{\left(n^{-1} \sum_{t=1}^{n} \hat{x}_{t i; J}^{\bot^2}\right)^{1/2}+\rho_{1}\left ( \frac{\log p_{n} }{n}  \right ) ^{1/2}}\right| \right.\nonumber \\ 
	&\left. + \max _{\sharp(J) \leq K_{n}-1, i,j \notin J}\left|\sum_{j \notin J} \beta_{j} \left\{ \frac{n^{-1} \sum_{t=1}^{n} x_{t j} \hat{x}_{t i; J}^{\perp}}{\left(n^{-1} \sum_{t=1}^{n} \hat{x}_{t i; J}^{\bot^2}\right)^{1/2}+\rho_{1}\left ( \frac{\log p_{n} }{n}  \right ) ^{1/2}} - \frac{E\left(x_{j} x_{i; J}^{\perp}\right)}{E\left(x_{i; J}^{\perp^2}\right)^{1/2}} \right\}\right|> 2\lambda^{*}\left(\frac{\log p_{n}}{n}\right)^{1/2}\right) \nonumber \\
   &\le P\left(\max _{\sharp(J) \leq K_{n}-1, i \notin J}\left|\frac{n^{-1} \sum_{t=1}^{n} \varepsilon_{t} \hat{x}_{t i; J}^{\perp}}{\left(n^{-1} \sum_{t=1}^{n} \hat{x}_{t i; J}^{\bot^2}\right)^{1/2}+\rho_{1}\left ( \frac{\log p_{n} }{n}  \right ) ^{1/2}}\right|>\lambda^{*}\left ( \frac{\log p_{n} }{n}  \right ) ^{1/2}\right) \nonumber \\&+
       P\left(\max _{\sharp(J) \leq K_{n}-1, i,j \notin J}\left|\sum_{j \notin J} \beta_{j} \left\{ \frac{n^{-1} \sum_{t=1}^{n} x_{t j} \hat{x}_{t i; J}^{\perp}}{\left(n^{-1} \sum_{t=1}^{n} \hat{x}_{t i; J}^{\bot^2}\right)^{1/2}+\rho_{1}\left ( \frac{\log p_{n} }{n}  \right ) ^{1/2}} - \frac{E\left(x_{j} x_{i; J}^{\perp}\right)}{E\left(x_{i; J}^{\perp^2}\right)^{1/2}} \right\}\right|>\lambda^{*}\left ( \frac{\log p_{n} }{n}  \right ) ^{1/2} \right) \ \nonumber\\
   &=o(1).\nonumber
\end{align}

For any  $0<\xi<1 $, let $\tilde{\xi}=2 /(1-\xi)$ and take 
\begin{equation}
	B_{n}(m)=\left\{\min _{0 \leq i \leq m-1} \max _{1 \leq j \leq p_{n}}\left|\mu_{\hat{J}_{i}, j}\right|>\tilde{\xi} s\left ( \frac{\log p_{n} }{n}  \right ) ^{1/2}\right\},
    \nonumber
\end{equation}
in which we set $\mu_{J, j}=0 $ if $ j \in J $ , and $\mu_{\hat{J}_{0}, j}=\mu_{\emptyset, j} $ . We now show that for all $1 \leq q \leq m$, 

\begin{equation}
	\left|\mu_{\hat{J}_{q-1}, \hat{j}_{q}}\right| \geq \xi \max _{1 \leq i \leq p_{n}}\left|\mu_{\hat{J}_{q-1}, i}\right| \text { on } A_{n}(m) \bigcap B_{n}(m),
	\label{ref13}
\end{equation}
Note that on $A_{n}(m) \bigcap B_{n}(m)$,
\begin{align}
\left|\mu_{\hat{J}_{q-1}, \hat{j}_{q}}\right| & \geq-\left|\hat{\mu}_{\hat{J}_{q-1}, \hat{j}_{q}}-\mu_{\hat{J}_{q-1}, \hat{j}_{q}}\right|+\left|\hat{\mu}_{\hat{J}_{q-1}, \hat{j}_{q}}\right| \nonumber \\
& \geq-\max _{(J, i): \sharp(J) \leq m-1, i \notin J}\left|\hat{\mu}_{J, i}-\mu_{J, i}\right|+\left|\hat{\mu}_{\hat{J}_{q-1}, \hat{j}_{q}}\right| \nonumber\\
& \geq-s\left ( \frac{\log p_{n} }{n}  \right ) ^{1/2}+\max _{1 \leq j \leq p_{n}}\left|\hat{\mu}_{\hat{J}_{q-1}, j}\right|\left(\text { since }\left|\hat{\mu}_{\hat{J}_{q-1}, \hat{j}_{q}}\right|=\max _{1 \leq j \leq p_{n}}\left|\hat{\mu}_{\hat{J}_{q-1}, j}\right|\right) \nonumber\\
& \geq-2 s\left ( \frac{\log p_{n} }{n}  \right ) ^{1/2}+\max _{1 \leq j \leq p_{n}}\left|\mu_{\hat{J}_{q-1}, j}\right| \label{ref34}\\
& \geq \xi\max_{1 \leq j \leq p_{n}}\left|\mu_{\hat{J}_{q-1}, j}\right|,\nonumber
\end{align}

since $2 s\left ( \frac{\log p_{n} }{n}  \right ) ^{1/2}<(2 / \tilde{\xi}) \max _{1 \leq j \leq p_{n}}\left|\mu_{\hat{J}_{q-1}, j}\right|$  on  $B_{n}(m)$ and $1-\xi=2 / \tilde{\xi}$.

Consider the semi-population version of the GSFR, which uses the variable selector \(\left(\hat{j}_{1}, \hat{j}_{2}, \dots \right)\) and approximates \(y(\mathbf{x})\) by \(y_{\hat{J}_{m}}(\mathbf{x})\). Equation (\ref{ref13}) demonstrates that this semi-population version of GSFR is consistent with the PrWGA in \citet{borodinProjectionGreedyAlgorithm2021} on \(A_{n}(m) \cap B_{n}(m)\). Using Lemma \ref{lemma4}, we prove

\begin{equation}
	E_{n}\left[\left\{y(\mathbf{x})-y_{\hat{J}_{m}}(\mathbf{x})\right\}^{2}\right] \leq\left(\sum_{j=1}^{p_{n}}\left|\beta_{j}\right|\right)^{2}\left(1+m \xi^{2}\right)^{-1} \text { on } A_{n}(m) \bigcap B_{n}(m) .
	\label{ref14}
\end{equation}

For  $0 \leq i \leq m-1, E_{n}\left[\left\{y(\mathbf{x})-y_{\hat{J}_{m}}(\mathbf{x})\right\}^{2}\right] \leq E_{n}\left[\left\{y(\mathbf{x})-y_{\hat{J}_{i}}(\mathbf{x})\right\}^{2}\right]$, and therefore 

\begin{align}
E_{n}\left[\left\{y(\mathbf{x})-y_{\hat{J}_{m}}(\mathbf{x})\right\}^{2}\right] & \leq \min _{0 \leq i \leq m-1} E_{n}\left\{\left(y(\mathbf{x})-y_{\hat{J}_{i}}(\mathbf{x})\right)\left(\sum_{j=1}^{p_{n}} \beta_{j} x_{j}\right)\right\} \nonumber\\
& \leq \min _{0 \leq i \leq m-1}  \max _{1 \leq j \leq p_{n}}\left |  \frac{E_{n}\left\{\left(y(\mathbf{x})-y_{\hat{J}_{i}}(\mathbf{x})\right)\left(x_{j}-x^{(\hat{J}_{i})}_{j} \right)\right\}}{{E_{n}\left(x_{j; \hat{J}_{i}}^{\perp^2}\right)}^{\frac{1}{2} } } \right | \sum_{j=1}^{p_{n}} \left |\beta_{j}  \right |     \nonumber\\
& = \min _{0 \leq i \leq m-1} \max _{1 \leq j \leq p_{n}}\left|\mu_{\hat{J}_{i}, j}\right| \sum_{j=1}^{p_{n}}\left|\beta_{j}\right| \nonumber\\
& \leq \tilde{\xi} s\left ( \frac{\log p_{n} }{n}  \right ) ^{1/2} \sum_{j=1}^{p_{n}}\left|\beta_{j}\right| \text { on } B_{n}^{c}(m) .\nonumber
\end{align}

Combining this with (\ref{assumption4}) and (\ref{ref14}), and the assumption that $m \leq K_{n}=O((n/\log p_{n})^{1 / 2})$ yields

\begin{equation}
	E_{n}\left[\left\{y(\mathbf{x})-y_{\hat{J}_{m}}(\mathbf{x})\right\}^{2}\right] I_{A_{n}(m)} \leq C^{*} m^{-1} 
	\label{ref16}
\end{equation}
for some $C^{*}>0$. In addition, since (\ref{ref9}) and (\ref{ref14}) $\forall \varepsilon >0$, $\exists M_{1} > C^{*}$, such that

\begin{equation*}
	 P\left (\max _{1 \leq m \leq K_{n}} m E_{n}\left[\left\{y(\mathbf{x})-y_{\hat{J}_{m}}(\mathbf{x})\right\}^{2}\right]\ge  M_{1}\right ) \le 1-P\left (A_{n}\left(K_{n}\right)  \right ) ,
\end{equation*}
and since $\lim_{n \to \infty} P\left (A_{n}\left(K_{n}\right)  \right )=1$ , $\exists N_{\varepsilon }$ , s.t. $1-P\left (A_{n}\left(K_{n}\right)  \right )<\varepsilon$ for $n>N_{\varepsilon }$. Take $M_{2\varepsilon}$ such that for $1\le n \le N_{\varepsilon }$
\begin{equation*}
	P\left (\max _{1 \leq m \leq K_{n}} m E_{n}\left[\left\{y(\mathbf{x})-y_{\hat{J}_{m}}(\mathbf{x})\right\}^{2}\right]\ge  M_{2\varepsilon}\right ) \le \varepsilon ,
\end{equation*}
taking $M_{3\varepsilon}=\max\left \{ M_{1\varepsilon}, M_{2\varepsilon} \right \}$ we have $\forall n >0$

\begin{equation*}
	P\left (\max _{1 \leq m \leq K_{n}} m E_{n}\left[\left\{y(\mathbf{x})-y_{\hat{J}_{m}}(\mathbf{x})\right\}^{2}\right]\ge  M_{3\varepsilon}\right ) \le \varepsilon,
\end{equation*}
following this we conclude that
\begin{equation}
	\max _{1 \leq m \leq K_{n}} m E_{n}\left[\left\{y(\mathbf{x})-y_{\hat{J}_{m}}(\mathbf{x})\right\}^{2}\right]=O_{p}(1).
	\label{result1}
\end{equation}

Let $\mathbf{q}(J)=E\left(y \mathbf{x}_{J}\right)$  and  $\mathbf{Q}(J)=n^{-1} \sum_{t=1}^{n}\left(y_{t}-\mathbf{x}_{t}^{\prime}(J) \boldsymbol{\Gamma}^{-1}(J)\right. \mathbf{q}(J)) \mathbf{x}_{t}(J)$.

Note 
\begin{align}
	\hat{y}_{m}(x)-y_{\hat{J}_{m}}(x)&=\mathbf{x}^{\prime}\left ( \hat{J}_{m} \right ) \left[\hat{\boldsymbol{\Gamma}}^{-1}\left(\hat{J}_{m}\right) \frac{\sum_{t=1}^{n}  \mathbf{x}_{t}\left(\hat{J}_{m}\right) y_{t}}{n}-\boldsymbol{\Gamma}^{-1}\left(\hat{J}_{m}\right) \boldsymbol{q}\left(\hat{J}_{m}\right)\right]
	   \nonumber \\
	&=\mathbf{x}^{\prime}\left ( \hat{J}_{m} \right )\hat{\boldsymbol{\Gamma}}^{-1}\left(\hat{J}_{m}\right) \left[\frac{\sum_{t=1}^{n}  \mathbf{x}_{t}\left(\hat{J}_{m}\right) y_{t}}{n}-\hat{\boldsymbol{\Gamma}}\left(\hat{J}_{m}\right)\boldsymbol{\Gamma}^{-1}\left(\hat{J}_{m}\right) \boldsymbol{q}\left(\hat{J}_{m}\right)\right]
	\nonumber \\
	&=\mathbf{x}^{\prime}\left ( \hat{J}_{m} \right )\hat{\boldsymbol{\Gamma}}^{-1}\left(\hat{J}_{m}\right) \frac{1}{n} \sum_{t=1}^{n} \left (  y_{t}- \mathbf{x}^{\prime}_{t}\left(\hat{J}_{m}\right)\boldsymbol{\Gamma}^{-1}\left(\hat{J}_{m}\right) \boldsymbol{q}\left(\hat{J}_{m}\right)\right )\mathbf{x}_{t}\left(\hat{J}_{m}\right) \nonumber \\
	&=\mathbf{x}^{\prime}\left ( \hat{J}_{m} \right )\hat{\boldsymbol{\Gamma}}^{-1}\left(\hat{J}_{m}\right)\mathbf{Q}(\hat{J}_{m}), \nonumber 
\end{align}
using this we show
\begin{align}
	E_{n}&\left(\hat{y}_{m}(\mathbf{x})-y_{\hat{J}_{m}}(\mathbf{x})\right)^{2}
	\nonumber
	\\&=\mathbf{Q}^{\top}\left(\hat{J}_{m}\right) \hat{\boldsymbol{\Gamma}}^{-1}\left(\hat{J}_{m}\right) \boldsymbol{\Gamma}\left(\hat{J}_{m}\right) \hat{\boldsymbol{\Gamma}}^{-1}\left(\hat{J}_{m}\right) \mathbf{Q}\left(\hat{J}_{m}\right) \nonumber \\
	&\leq\left\|\mathbf{Q}\left(\hat{J}_{m}\right)\right\|^{2}\left\|\hat{\boldsymbol{\Gamma}}^{-1}\left(\hat{J}_{m}\right)\right\|^{2}\left\|\hat{\boldsymbol{\Gamma}}\left(\hat{J}_{m}\right)-\boldsymbol{\Gamma}\left(\hat{J}_{m}\right)\right\|+\left\|\mathbf{Q}\left(\hat{J}_{m}\right)\right\|^{2}\left\|\hat{\boldsymbol{\Gamma}}^{-1}\left(\hat{J}_{m}\right)\right\| ,\nonumber
\end{align}
where for $\mathbf{A} \in \mathbf{R}  ^{m\times n}$, $\left \|  \mathbf{A}  \right \|=\max_{\left \|  x  \right \|_{2}=1}{\left \|  \mathbf{A}x  \right \|}_{2} $.

To estimate $\left\|\mathbf{Q}\left(\hat{J}_{m}\right)\right\|^2$, we have
\begin{align}
\mathbf{Q}(\hat{J}_{m})
        &=\frac{1}{n} \sum_{t=1}^{n} y_{t}\mathbf{x}_{t}\left(\hat{J}_{m}\right)- \mathbf{x}^{\prime}_{t}\left(\hat{J}_{m}\right)\boldsymbol{\Gamma}^{-1}\left(\hat{J}_{m}\right) \boldsymbol{q}\left(\hat{J}_{m}\right)\mathbf{x}_{t}\left(\hat{J}_{m}\right)
           \nonumber \\
        &=\frac{1}{n} \sum_{t=1}^{n} \left ( \mathbf{x}_{t}^{\prime}\mathbf{\beta}+\varepsilon_{t} \right ) \mathbf{x}_{t}\left(\hat{J}_{m}\right)- \mathbf{x}^{\prime}_{t}\left(\hat{J}_{m}\right)\boldsymbol{\Gamma}^{-1}\left(\hat{J}_{m}\right) E\left ( \sum_{i=1}^{p_{n} }\left ( x_{i} \beta_{i}  \right )\mathbf{x}\left(\hat{J}_{m}\right)  \right ) \mathbf{x}_{t}\left(\hat{J}_{m}\right)
        \nonumber \\
        &=\frac{1}{n} \sum_{t=1}^{n}\varepsilon_{t} \mathbf{x}_{t}\left(\hat{J}_{m}\right)- \sum_{i=1}^{p_{n} }\beta_{i} \sum_{t=1}^{n}\left ( x_{ti}- \mathbf{x}^{\prime}_{t}\left(\hat{J}_{m}\right)\boldsymbol{\Gamma}^{-1}\left(\hat{J}_{m}\right)\mathbf{g}_{i} \left ( \hat{J}_{m} \right )  \right ) \mathbf{x}_{t}\left(\hat{J}_{m}\right)
        \label{ref15}.
\end{align}

Since $\forall j^{*} \in \hat{J}_{m}$
\begin{align}
	&\left | \frac{1}{n} \sum_{t=1}^{n} x_{ti}x_{tj^{*}}-\sigma _{ij^{*}}+\sigma _{ij^{*}} - \mathbf{g}^{\prime}_{i} \left ( \hat{J}_{m} \right )\boldsymbol{\Gamma}^{-1}\left(\hat{J}_{m}\right)\frac{1}{n} \sum_{t=1}^{n}\mathbf{x}_{t}\left(\hat{J}_{m}\right)x_{tj^{*}} \right | 
	\nonumber \\
	&=\left | \frac{1}{n} \sum_{t=1}^{n} x_{ti}x_{tj^{*}}-\sigma _{ij^{*}}+\mathbf{g}^{\prime}_{i} \left ( \hat{J}_{m} \right )\boldsymbol{\Gamma}^{-1}\left(\hat{J}_{m}\right)\mathbf{g}_{j^{*}} \left ( \hat{J}_{m} \right ) - \mathbf{g}^{\prime}_{i} \left ( \hat{J}_{m} \right )\boldsymbol{\Gamma}^{-1}\left(\hat{J}_{m}\right)\frac{1}{n} \sum_{t=1}^{n}\mathbf{x}_{t}\left(\hat{J}_{m}\right)x_{tj^{*}} \right |
	\nonumber \\
	&\le \max _{1\leq i, j \leq p_{n}}\left |n^{-1} \sum_{t=1}^{n} x_{t i} x_{t j}-\sigma_{i j} \right | \left ( 1+\max _{1 \leq \sharp(J) \leq K_{n}, 1 \leq l \leq p_{n}}\left\|\boldsymbol{\Gamma}^{-1}(J) \mathbf{g}_{l}(J)\right\|_{1} \right ) 
	\nonumber
\end{align}
combining (\ref{ref15}) we have
\begin{equation}
	\begin{aligned}
\left\|\mathbf{Q}\left(\hat{J}_{m}\right)\right\|^{2} \leq & 2 m \max _{1 \leq i \leq p_{n}}\left(n^{-1} \sum_{t=1}^{n} \varepsilon_{t} x_{t i}\right)^{2}+2 m \max _{1 \leq i, j \leq p_{n}}\left(n^{-1} \sum_{t=1}^{n} x_{t i} x_{t j}-\sigma_{i j}\right)^{2} \\
& \times\left(\sum_{j=1}^{p_{n}}\left|\beta_{j}\right|\right)^{2}\left(1+\max _{1 \leq \sharp(J) \leq K_{n}, 1 \leq l \leq p_{n}}\left\|\boldsymbol{\Gamma}^{-1}(J) \mathbf{g}_{l}(J)\right\|_{1}\right)^{2} .
\end{aligned}
\nonumber
\end{equation}

Combine this with (\ref{assumption5_equation2}), (\ref{assumption4}), Lemma \ref{lemma1} and Lemma \ref{lemma2}, we now prove
\begin{equation}
	\max _{1 \leq m \leq K_{n}}\left\{\frac{n\left\|\mathbf{Q}\left(\hat{J}_{m}\right)\right\|^{2}}{m \log p_{n}}\right\}=O_{p}(1).
	\label{result2}
\end{equation}

For every \(\varepsilon > 0\), take $M^{\prime}=\max\left \{ \frac{8M^{\prime}_{1}}{C}, \frac{8M^{\prime}_{1}}{\eta}  \right \} $, where  
\[
M^{\prime}_{1}=\left(\sum_{j=1}^{p_{n}}\left|\beta_{j}\right|\right)^{2}\left(1+\max _{1 \leq \sharp(J) \leq K_{n}, 1 \leq l \leq p_{n}}\left\|\boldsymbol{\Gamma}^{-1}(J) \mathbf{g}_{l}(J)\right\|_{1}\right)^{2},
\]
then \(\exists N^{\prime}_{1}, \forall n> N^{\prime}_{1},\) 

\begin{align*}
    P\Bigg( &\max_{1 \leq m \leq K_{n}} \left\{ \frac{n \left\| \mathbf{Q}\left(\hat{J}_{m}\right) \right\|^{2}}{m \log p_{n}} \right\} > M^{\prime} \Bigg) \\
    &\le P\Bigg( \max_{1 \leq m \leq K_{n}} \left\{ \frac{n \left\| \mathbf{Q}\left(\hat{J}_{m}\right) \right\|^{2}}{m \log p_{n}} \right\} > M^{\prime}, \Omega_{n} \Bigg) + \frac{\varepsilon}{2},
\end{align*}

\begin{align*}
    P\Bigg( &\max_{1 \leq m \leq K_{n}} \left\{ \frac{n \left\| \mathbf{Q}\left(\hat{J}_{m}\right) \right\|^{2}}{m \log p_{n}} \right\} > M^{\prime}, \Omega_{n} \Bigg) \\
    &\le P\left( \max_{1 \leq i \leq p_{n}} \left( \frac{\sum_{t=1}^{n} \varepsilon_{t} x_{t i}}{n} \right)^2 > \frac{M^{\prime}}{4} \frac{\log p_{n}}{n}, \Omega_{n} \right) \\
    &\quad + P\left( M^{\prime}_{1} \max_{1 \leq i,j \leq p_{n}} \left( \frac{\sum_{t=1}^{n} x_{t i} x_{t j}}{n} - \sigma_{i j} \right)^2 > \frac{M^{\prime}}{4} \frac{\log p_{n}}{n} \right).
\end{align*}

By Lemma \ref{lemma1}, $\exists N^{\prime}_{2} >0, \forall n>N^{\prime}_{2}$ such that

\begin{align*}
	P\Bigg(M^{\prime}_{1}&\max _{1 \leq i, j \leq p_{n}}\left(\frac{\sum_{t=1}^{n}x_{ti}x_{tj} }{n} -\sigma_{i j}\right)^2>\frac{M^{\prime}}{4}  \frac{\log{p_{n}}}{n}\ \Bigg) \\
        & \le p^2_{n} \max _{1 \leq i, j \leq p_{n}}P\left(\left | \frac{\sum_{t=1}^{n}x_{ti}x_{tj} }{n} -\sigma_{i j}\right |>(\frac{M^{\prime}}{4M^{\prime}_{1}}  \frac{\log{p_{n}}}{n})^{\frac{1}{2}} \right) \nonumber \\
	&\le \exp \left ( \left ( 2 -\frac{CM^{\prime}}{4M^{\prime}_{1}} \right )  \log{p_{n} }\right  )		\nonumber \\
	&< \frac{\varepsilon}{4}.
        \nonumber
\end{align*}

Similarly, by Lemma \ref{lemma2} we have $\exists N^{\prime}_{3} >0, \forall n>N^{\prime}_{3}$
\begin{equation*}
	P\left(\max _{1 \leq i \leq p_{n}} \left(\frac{ \sum_{t=1}^{n} \varepsilon_{t} x_{t i}}{n}\right)^2 > \frac{M^{\prime}}{4}  \frac{\log{p_{n}}}{n} , \Omega_{n}\right) \le p_{n}\exp \left ( -\eta\frac{M^{\prime}}{4M^{\prime}_{1}}  \log{p_{n}}\right ) < \frac{\varepsilon}{4}.
\end{equation*}

Take $N^{\prime}=\max\left \{N^{\prime}_{1} , N^{\prime}_{2},  N^{\prime}_{3} \right \} $, then $\forall n>N^{\prime}$ 

\begin{equation*}
	P\left (\max _{1 \leq m \leq K_{n}}\left\{\frac{n\left\|\mathbf{Q}\left(\hat{J}_{m}\right)\right\|^{2}}{m \log p_{n}}\right\}>M^{\prime}  \right )< \varepsilon.
\end{equation*}

Take $M^{\prime}_{2\varepsilon }$ such that $\forall n \le N^{\prime}$
\begin{equation*}
	P\left (\max _{1 \leq m \leq K_{n}}\left\{\frac{n\left\|\mathbf{Q}\left(\hat{J}_{m}\right)\right\|^{2}}{m \log p_{n}}\right\}>M^{\prime}_{2\varepsilon }  \right )< \varepsilon,
\end{equation*}
and take $M^{*}_{\varepsilon }=\max\left \{M^{\prime}_{\varepsilon },   M^{\prime}_{2\varepsilon }\right \} $, then $\forall n>0$

\begin{equation*}
	P\left (\max _{1 \leq m \leq K_{n}}\left\{\frac{n\left\|\mathbf{Q}\left(\hat{J}_{m}\right)\right\|^{2}}{m \log p_{n}}\right\}>M^{*}_{\varepsilon }  \right )< \varepsilon,
\end{equation*}
result (\ref{result2}) follows immediately.
\\
Moreover, by Lemma \ref{lemma1} and take $M^{\prime}_{3}>(\frac{2}{c})^{\frac{1}{2}}+\frac{1}{\delta}$,  
$\forall \varepsilon, \exists N^{\prime}_{4\varepsilon} >0, \forall n>N^{\prime}_{4\varepsilon}$ 
\begin{align*}
P\Bigg(&\max _{1 \leqslant m \leqslant k_{n}}\left\|\hat{\mathbf{\Gamma} }^{-1}\left(\hat{J}_{m}\right)\right\|>M^{\prime}_{3} \Bigg)  \\
&\leqslant P\left(\max _{1 \leqslant m \leqslant k_{n}}\left\|\hat{\mathbf{\Gamma}}^{-1}\left(\hat{J}_{m}\right)-\mathbf{\Gamma}^{-1}\left(\hat{J}_{m}\right)\right\|+\max _{1 \leqslant m \leqslant k_{n} }\left\|\mathbf{\Gamma}^{-1}\left(\hat{J}_{m}\right)\right\|>M^{\prime}_{3 }\right) \nonumber \\
&\leqslant P\left(\max _{1 \leqslant \sharp(J) \leqslant k_{n}}\left\|\hat{\mathbf{\Gamma}}^{-1}(J)-\mathbf{\Gamma}^{-1}(J)\right\|+\max _{1 \leqslant \sharp(J) \leqslant k_{n}}\left\|\Gamma^{-1}(J)\right\|>M^{\prime}_{3 }\right) \nonumber\\
&\leqslant P\left(\max _{1 \leqslant \sharp(J) \leqslant k_{n}}\left\|\hat{\mathbf{\Gamma}}^{-1}(J)-\mathbf{\Gamma}^{-1}(J)\right\|>M^{\prime}_{3}-\frac{1}{\delta}\right),\nonumber\\
&\le \nonumber \varepsilon.
\end{align*}

Take $M^{\prime}_{4\varepsilon }$ such that $\forall n \le N^{\prime}_{4\varepsilon}$
\begin{equation*}
	P\left(\max _{1 \leqslant m \leqslant k_{n}}\left\|\hat{\mathbf{\Gamma} }^{-1}\left(\hat{J}_{m}\right)\right\|>M^{\prime}_{4\varepsilon } \right)\le \nonumber \varepsilon,
\end{equation*}
we similarly conclude that 
\begin{equation}
	\max _{1 \leq m \leq K_{n}}\left\|\hat{\boldsymbol{\Gamma}}^{-1}\left(\hat{J}_{m}\right)\right\|=O_{p}(1), \max _{1 \leq m \leq K_{n}}\left\|\hat{\boldsymbol{\Gamma}}\left(\hat{J}_{m}\right)-\boldsymbol{\Gamma}\left(\hat{J}_{m}\right)\right\|=O_{p}(1),
	\nonumber
\end{equation}
and
\begin{equation}
	\max _{1 \leq m \leq K_{n}} \frac{n E_{n}\left[\left\{\hat{y}_{m}(\mathbf{x})-y_{\hat{J}_{m}}(\mathbf{x})\right\}^{2}\right]}{m \log p_{n}}=O_{p}(1)
	\label{result3}.
\end{equation}

By combining the expressions
\[
E_{n}\left[\left\{y(\mathbf{x})-\hat{y}_{m}(\mathbf{x})\right\}^{2}\right] = E_{n}\left[\left\{y(\mathbf{x})-y_{\hat{J}_{m}}(\mathbf{x})\right\}^{2}\right] + E_{n}\left[\left\{\hat{y}_{m}(\mathbf{x})-y_{\hat{J}_{m}}(\mathbf{x})\right\}^{2}\right],
\]
and incorporating the results from (\ref{result1}) and (\ref{result3}), we conclude that:

\begin{equation}
	\\max _{1 \leq m \leq K_{n}}\left(\frac{E\left[\left\{y(\mathbf{x})-\hat{y}_{m}(\mathbf{x})\right\}^{2} \mid y_{1}, \mathbf{x}_{1}, \ldots, y_{n}, \mathbf{x}_{n}\right]}{m^{-1}+n^{-1} m \log p_{n}}\right)=O_{p}(1).
    \nonumber
\end{equation}

\end{proof}

\section{Proof of Theorem \ref{theorem2}}
\begin{proof}
The structure of this proof parallels that of \citet{ingStepwiseRegressionMethod2011}, with appropriate adjustments for GSFR. Let \(a > 0\) and define \(A_n(m)\) by (\ref{ref9}), where \(m = \lfloor a n^{\gamma} \rfloor = o(K_n)\). By (\ref{ref9}) and (\ref{ref16}), we have
\[
\lim_{n \to \infty} P(A_n^c(m)) \leq \lim_{n \to \infty} P(A_n^c(K_n)) = 0,
\]
and
\[
E_n\{[y(\mathbf{x}) - y_{\hat{J}_m}(\mathbf{x})]^2 \} I_{A_n(m)} \leq C^* m^{-1}.
\]

From this, it follows that
\begin{equation}
	\lim_{n \to \infty} P(F_n) = 0, \quad \text{where } F_n = \left\{ E_n[y(\mathbf{x}) - y_{\hat{J}_m}(\mathbf{x})]^2 > C^* m^{-1} \right\}.
	\label{ref20}
\end{equation}

For \(J \subseteq \{1, \dots, p_n\}\) and \(j \in J\), let \(\hat{\beta}_j(J)\) denote the coefficient of \(x_j\) in the best linear predictor \(\sum_{i \in J} \hat{\beta}_i(J)x_i\) of \(y\) that minimizes \(E[y - \sum_{i \in J} \lambda_i x_i]^2\). Define \(\hat{\beta}_j(J) = 0\) if \(j \notin J\). Note that
\begin{equation}
	E_n[y(\mathbf{x}) - y_{\hat{J}_m}(\mathbf{x})]^2 = E_n\left\{\sum_{j \in \hat{J}_m \cup N_n} (\beta_j - \hat{\beta}_j(\hat{J}_m)) x_j \right\}^2.
	\label{ref17}
\end{equation}

From (\ref{assumption4}) and (\ref{assumption6_equation2}), it follows that
\[
\sharp(N_n) = O(n^{\gamma/2}),
\]
yielding \(\sharp(\hat{J}_m \cup N_n) = o(K_n)\). It then follows from (\ref{ref17}) that for all large \(n\),
\begin{equation}
	E_n\{[y(\mathbf{x}) - y_{\hat{J}_m}(\mathbf{x})]^2\} \geq \left(\min_{j \in N_n} \beta_j^2\right) \min_{1 \leq \sharp(J) \leq K_n} \lambda_{\min}(\Gamma(J)),
	\label{ref18}
\end{equation}

on \(\{N_n \cap \hat{J}_m^c \neq \emptyset\}\).

Combining this result with (\ref{ref18}), (\ref{assumption6_equation2}), and (\ref{assumption5_equation1}), we obtain
\begin{equation}
	E_n\{[y(\mathbf{x}) - y_{\hat{J}_m}(\mathbf{x})]^2\} \geq b n^{-\gamma},
	\label{ref19}
    \nonumber
\end{equation}
for some \(b > 0\) and all large \(n\). By choosing \(a\) in \(m = \lfloor a n^\gamma \rfloor\) large enough, we have \(b n^{-\gamma} > C^* m^{-1}\), implying that \(\{N_n \cap \hat{J}_m^c \neq \emptyset\} \subseteq F_n\), where \(F_n\) is defined above. Hence, by (\ref{ref20}),
\[
\lim_{n \to \infty} P(N_n \subseteq \hat{J}_m) = 1.
\]

Therefore, the GSFR algorithm path, which terminates after \( m = \lfloor a n^\gamma \rfloor \) iterations, includes all relevant regressors with probability approaching 1. This result also holds for the GSFR path terminating after \( K_n \) iterations, provided that \( K_n / m \to \infty \).
\end{proof}

\section{Proof of Theorem \ref{theorem3}}
\begin{proof}

First, from \eqref{ref9} and Theorem \ref{theorem2}, we know the probability of the event \( A_{n}\left(K_{n}\right) \) and \( A^{\prime}_{n}\left(K_{n}\right) \) satisfies
\begin{align*}
    \lim_{n \to \infty} P\left(A_{n}\left(K_{n}\right)\right) = 1, \\
     \lim_{n \to \infty} P\left(A^{\prime}_{n}\left(K_{n}\right)\right) = 1,
    \nonumber
\end{align*}
where
\begin{align*}
    &A_{n}(m) = \left\{\max_{(J, i): \sharp(J) \leq m-1, i \notin J} \left|\hat{\mu}_{J, i} - \mu_{J, i}\right| \leq s\left(\frac{\log p_{n}}{n}\right)^{1/2}\right\},\\
    &A^{\prime}_{n}(m) = \left\{\tilde{k}_n<K_{n}\right\}.
    \nonumber
\end{align*}

By definition,  on the event \(  A_{n}\left(K_{n}\right)\cap  A^{\prime}_{n}\left(K_{n}\right) \), we have
\begin{equation}
    0 < \hat{\triangle}_{\hat{J}_{\hat{k}_n}} \le \hat{\triangle}_{\hat{J}_{\tilde{k}_n}}.
    \label{ref25}
\end{equation}

 Using \eqref{ref25}, we have
\begin{equation}
    \hat{\triangle}_{\hat{J}_{\hat{k}_n}} = \frac{\left| \hat{\mu}_{\hat{J}_{\hat{k}_n}, \hat{j}_{\hat{k}_n+1}} \right| + \rho_2 \frac{1}{n^{1.5\gamma + \varepsilon_0 }}}{\left| \hat{\mu}_{\hat{J}_{{\hat{k}_n}-1}, \hat{j}_{{\hat{k}_n}}} \right| + \rho_2 \frac{1}{n^{1.5\gamma + \varepsilon_0 }}} 
    \le \hat{\triangle}_{\hat{J}_{\tilde{k}_n}} = \frac{\left| \hat{\mu}_{\hat{J}_{{\tilde{k}_n}}, \hat{j}_{{\tilde{k}_n}+1}} \right| + \rho_2 \frac{1}{n^{1.5\gamma + \varepsilon_0 }}}{\left| \hat{\mu}_{\hat{J}_{{\tilde{k}_n}-1}, \hat{j}_{{\tilde{k}_n}}} \right| + \rho_2 \frac{1}{n^{1.5\gamma + \varepsilon_0}}}.
    \label{ref26}
    \nonumber
\end{equation}

From \eqref{ref9}, we obtain
\begin{equation}
    \frac{\left| \mu_{\hat{J}_{\hat{k}_n}, \hat{j}_{\hat{k}_n+1}} \right| + \rho_2\frac{1}{n^{1.5\gamma + \varepsilon_0 }} - s \left( \frac{\log p_n}{n} \right)^{1/2}}{\left| \mu_{\hat{J}_{{\hat{k}_n}-1}, \hat{j}_{{\hat{k}_n}}} \right| + \rho_2\frac{1}{n^{1.5\gamma + \varepsilon_0 }} + s \left( \frac{\log p_n}{n} \right)^{1/2}} 
    \le \frac{\left| \mu_{\hat{J}_{{\tilde{k}_n}}, \hat{j}_{{\tilde{k}_n}+1}} \right| + \rho_2\frac{1}{n^{1.5\gamma + \varepsilon_0 }} + s \left( \frac{\log p_n}{n} \right)^{1/2}}{\left| \mu_{\hat{J}_{{\tilde{k}_n}-1}, \hat{j}_{{\tilde{k}_n}}} \right| + \rho_2\frac{1}{n^{1.5\gamma + \varepsilon_0 }} - s \left( \frac{\log p_n}{n} \right)^{1/2}}.
    \label{ref27}
\end{equation}

It is easy to see the following relationship: for \( m \ge \tilde{k}_n \), we have \( \left| \mu_{\hat{J}_{m}, \hat{j}_{m+1}} \right| = 0 \). Thus, on the event \( \{\hat{k}_n > \tilde{k}_n\} \cap A_{n}\left(K_{n}\right)\cap  A^{\prime}_{n}\left(K_{n}\right) \), it follows that 
\(\left| \mu_{\hat{J}_{\hat{k}_n-1}, \hat{j}_{\hat{k}_n}} \right| = 0\), \(\left| \mu_{\hat{J}_{\hat{k}_n}, \hat{j}_{\hat{k}_n+1}} \right| = 0\) and \(\left| \mu_{\hat{J}_{\tilde{k}_n}, \hat{j}_{\tilde{k}_n+1}} \right| = 0\). 

Consequently, we obtain the bounds
\begin{equation}
    \frac{\rho_2\frac{1}{n^{1.5\gamma + \varepsilon_0 }} - s \left( \frac{\log p_n}{n} \right)^{1/2}}{\rho_2\frac{1}{n^{1.5\gamma + \varepsilon_0}} + s \left( \frac{\log p_n}{n} \right)^{1/2}} \leq \hat{\triangle}_{\hat{J}_{\hat{k}_n}} 
    \leq \frac{\rho_2\frac{1}{n^{1.5\gamma + \varepsilon_0 }} + s \left( \frac{\log p_n}{n} \right)^{1/2}}{\left| \mu_{\hat{J}_{\tilde{k}_n-1}, \hat{j}_{\tilde{k}_n}} \right| + \rho_2\frac{1}{n^{1.5\gamma + \varepsilon_0 }} - s \left( \frac{\log p_n}{n} \right)^{1/2}}.
    \label{ref28}
\end{equation}

Now, because \(N_{n} \cap \hat{J}_{\tilde{k}_n-1}^{c} \neq \emptyset\). From \eqref{ref18}, we have
\begin{equation}
    E_n\left\{\left[y(\mathbf{x}) - y_{\hat{J}_{\tilde{k}_n-1}}(\mathbf{x})\right]^2\right\} 
    \geq \left(\min_{j \in N_n} \beta_j^2\right) \min_{1 \leq \sharp(J) \leq K_n} \lambda_{\min}(\Gamma(J)) \geq b n^{-\gamma},
    \label{ref29}
\end{equation}
for some \(b > 0\). 

Since \(\left| \mu_{\hat{J}_{\tilde{k}_n-1}, \hat{j}_{\tilde{k}_n}} \right| = \sqrt{\sigma_{\hat{J}_{\tilde{k}_n-1}}^2 - \sigma_{\hat{J}_{\tilde{k}_n}}^2}\), and \(\sigma_{\hat{J}_{\tilde{k}_n}}^2 = 0\) (as all relevant variables are already selected), \eqref{ref29} implies
\begin{equation}
    \left| \mu_{\hat{J}_{\tilde{k}_n-1}, \hat{j}_{\tilde{k}_n}} \right| = E_n\left\{\left[y(\mathbf{x}) - y_{\hat{J}_{\tilde{k}_n-1}}(\mathbf{x})\right]^2\right\}^{1/2} 
    \geq \sqrt{b} n^{-\frac{1}{2}\gamma}.
    \label{ref30}
\end{equation}

Combining \eqref{ref28} and \eqref{ref30}, we have
\begin{equation}
        \frac{\rho_2\frac{1}{n^{1.5\gamma + \varepsilon_0}} - s \left( \frac{\log p_n}{n} \right)^{1/2}}{\rho_2\frac{1}{n^{1.5\gamma + \varepsilon_0 }} + s \left( \frac{\log p_n}{n} \right)^{1/2}} \leq \hat{\triangle}_{\hat{J}_{\hat{k}_n}} 
    \leq \frac{\rho_2\frac{1}{n^{1.5\gamma + \varepsilon_0 }} + s \left( \frac{\log p_n}{n} \right)^{1/2}}{\left ( \rho_2 + \sqrt{b} \right ) \frac{1}{n^{1.5\gamma + \varepsilon_0 }} - s \left( \frac{\log p_n}{n} \right)^{1/2}}.
    \label{ref31}
    \nonumber
\end{equation}

Since \(K_n / n^{1.5\gamma + \varepsilon_0} \to \infty\), for sufficiently large \(n\), the following inequalities hold:
\[
1 - \varepsilon_1 \leq  \frac{\rho_2\frac{1}{n^{1.5\gamma + \varepsilon_0}} - s \left( \frac{\log p_n}{n} \right)^{1/2}}{\rho_2\frac{1}{n^{1.5\gamma + \varepsilon_0 }} + s \left( \frac{\log p_n}{n} \right)^{1/2}} ,
\]
and
\[
\frac{\rho_2\frac{1}{n^{1.5\gamma + \varepsilon_0 }} + s \left( \frac{\log p_n}{n} \right)^{1/2}}{\left ( \rho_2 + \sqrt{b} \right ) \frac{1}{n^{1.5\gamma + \varepsilon_0 }} - s \left( \frac{\log p_n}{n} \right)^{1/2}}
\leq \frac{\rho_2}{\rho_2 + \sqrt{b}} + \varepsilon_1,
\]
where \(0 < \varepsilon_1 < \frac{\sqrt{b}}{2(\rho_2 + \sqrt{b})}\).

Thus, on \(\{\hat{k}_n > \tilde{k}_n\} \cap A_{n}\left(K_{n}\right)\) with sufficiently large n, we have
\begin{equation}
   1 - \varepsilon_1
     \leq \hat{\triangle}_{\hat{J}_{\hat{k}_n}} 
     \leq \frac{\rho_2}{\rho_2 + \sqrt{b}} + \varepsilon_1,
    \label{ref31}
    \nonumber
\end{equation}
which leads to a contradiction. Therefore,
\begin{align}
    &\lim_{n \to \infty} P(\hat{k}_n > \tilde{k}_n,A_{n}\left(K_{n}\right)\cap A^{\prime}_{n}\left(K_{n}\right) ) = 0, \nonumber \\
     & \lim_{n \to \infty} P(\hat{k}_n > \tilde{k}_n) = 0.
    \label{ref32}
\end{align}

On the event \(\{\hat{k}_n < \tilde{k}_n\} \cap A_{n}\left(K_{n}\right)\cap A^{\prime}_{n}\left(K_{n}\right)\), using \eqref{assumption4}, we have 
\begin{equation}
    0 \leq \left| \mu_{\hat{J}_{\hat{k}_n-1}, \hat{j}_{\hat{k}_n}} \right| \leq E\left[y\left(\mathbf{x}\right)^2\right]^{1/2} \leq M,
    \nonumber
\end{equation}
for some \(M > 0\).

Combining this with \eqref{ref27}, we obtain
\begin{equation}
    \frac{\left| \mu_{\hat{J}_{\hat{k}_n}, \hat{j}_{\hat{k}_n+1}} \right| + \rho_2\frac{1}{n^{1.5\gamma + \varepsilon_0 }} - s \left( \frac{\log p_n}{n} \right)^{1/2}}{M + \rho_2\frac{1}{n^{1.5\gamma + \varepsilon_0 }} + s \left( \frac{\log p_n}{n} \right)^{1/2}} 
    \leq \frac{\rho_2\frac{1}{n^{1.5\gamma + \varepsilon_0 }} + s \left( \frac{\log p_n}{n} \right)^{1/2}}{\left| \mu_{\hat{J}_{{\tilde{k}_n}-1}, \hat{j}_{{\tilde{k}_n}}} \right| + \rho_2\frac{1}{n^{1.5\gamma + \varepsilon_0 }} - s \left( \frac{\log p_n}{n} \right)^{1/2}}.
    \label{ref33}
\end{equation}

From \eqref{ref34}, we note that
\begin{equation}
    -2 s\left( \frac{\log p_{n} }{n} \right)^{1/2} + \max _{1 \leq j \leq p_{n}} \left|\mu_{\hat{J}_{\hat{k}_n}, j}\right| 
    \leq \left| \mu_{\hat{J}_{\hat{k}_n}, \hat{j}_{\hat{k}_n+1}} \right|.
    \label{ref35}
\end{equation}

Moreover,  we have
\begin{align}
    E_{n}\left[\left\{y(\mathbf{x}) - y_{\hat{J}_{\hat{k}_n}}(\mathbf{x})\right\}^{2}\right] 
    &= E_{n}\left\{\left(y(\mathbf{x}) - y_{\hat{J}_{\hat{k}_n}}(\mathbf{x})\right) \left(\sum_{j = 1}^{p_{n}} \beta_{j} x_{j}\right)\right\} \nonumber \\
    &\leq \max _{1 \leq j \leq p_{n}} \left|\mu_{\hat{J}_{\hat{k}_n}, j}\right| \sum_{j = 1}^{p_{n}} \left|\beta_{j}\right| \nonumber \\
    &\leq M \max _{1 \leq j \leq p_{n}} \left|\mu_{\hat{J}_{\hat{k}_n}, j}\right|.
    \label{ref36}
\end{align}

Using \eqref{ref35} and \eqref{ref36}, we obtain
\begin{equation}
    \frac{E_{n}\left[\left\{y(\mathbf{x}) - y_{\hat{J}_{\hat{k}_n}}(\mathbf{x})\right\}^{2}\right]}{M} 
    \leq 2 s\left( \frac{\log p_{n} }{n} \right)^{1/2} + \left| \mu_{\hat{J}_{\hat{k}_n}, \hat{j}_{\hat{k}_n+1}} \right|.
    \label{ref37}
\end{equation}

Combining \eqref{ref33}, \eqref{ref37}, and \eqref{ref18}, and noting that \(K_n / n^{1.5\gamma + \varepsilon_0} \to \infty\), for sufficiently large n we deduce
\begin{align}
    \left| \mu_{\hat{J}_{{\tilde{k}_n}-1}, \hat{j}_{{\tilde{k}_n}}} \right| &+ \rho_2\frac{1}{n^{1.5\gamma + \varepsilon_0 }} - s \left( \frac{\log p_n}{n} \right)^{1/2}\\
    &\leq \frac{\left[M + \rho_2\frac{1}{n^{1.5\gamma + \varepsilon_0 }} + s \left( \frac{\log p_n}{n} \right)^{1/2}\right] [\rho_2\frac{1}{n^{1.5\gamma + \varepsilon_0 }} + s \left( \frac{\log p_n}{n} \right)^{1/2}]}{\left| \mu_{\hat{J}_{{\hat{k}_n}}, \hat{j}_{{\hat{k}_n}+1}} \right| + \rho_2\frac{1}{n^{1.5\gamma + \varepsilon_0 }} - s \left( \frac{\log p_n}{n} \right)^{1/2}} \nonumber \\
    &= \frac{\left[M + \rho_2\frac{1}{n^{1.5\gamma + \varepsilon_0 }} + s \left( \frac{\log p_n}{n} \right)^{1/2}\right] [\rho_2\frac{1}{n^{1.5\gamma + \varepsilon_0 }} + s \left( \frac{\log p_n}{n} \right)^{1/2}]}{\left| \mu_{\hat{J}_{{\hat{k}_n}}, \hat{j}_{{\hat{k}_n}+1}} \right| + 2s \left( \frac{\log p_n}{n} \right)^{1/2} + \rho_2\frac{1}{n^{1.5\gamma + \varepsilon_0 }} - 3s \left( \frac{\log p_n}{n} \right)^{1/2}} \nonumber \\
    &\leq \frac{\left[M + \rho_2\frac{1}{n^{1.5\gamma + \varepsilon_0 }} + s \left( \frac{\log p_n}{n} \right)^{1/2}\right] [\rho_2\frac{1}{n^{1.5\gamma + \varepsilon_0 }} + s \left( \frac{\log p_n}{n} \right)^{1/2}]}{\frac{b}{M} n^{-\gamma} + \rho_2\frac{1}{n^{1.5\gamma + \varepsilon_0 }} - 3s \left( \frac{\log p_n}{n} \right)^{1/2}} \nonumber \\
     &\leq \frac{2\left ( M+\rho_2 + s \right )  \rho_2\frac{1}{n^{1.5\gamma + \varepsilon_0 }}}{\frac{b}{M} n^{-\gamma} + \rho_2\frac{1}{n^{1.5\gamma + \varepsilon_0 }} - 3s \left( \frac{\log p_n}{n} \right)^{1/2}} \nonumber \\
    &\leq \frac{2\left ( M+\rho_2 + s \right )  \rho_2\frac{1}{n^{1.5\gamma + \varepsilon_0 }}}{\frac{b}{M} n^{-\gamma} } \nonumber \\
    &\leq 2\frac{M\left ( M+\rho_2 + s \right )  \rho_2}{b}\frac{1}{n^{\frac{1}{2} \gamma + \varepsilon_0 }}.
    \label{ref38}
\end{align}

From \eqref{ref38}, we deduce
\begin{equation}
    \left| \mu_{\hat{J}_{{\tilde{k}_n}-1}, \hat{j}_{{\tilde{k}_n}}} \right| \leq 2\frac{M\left ( M+\rho_2 + s \right )  \rho_2}{b}n^{-\left ( \frac{1}{2} \gamma + \varepsilon_0 \right )  },
    \nonumber
\end{equation}
which contradicts \eqref{ref30}. Therefore, we conclude
\begin{align}
   &\lim_{n \to \infty} P(\hat{k}_n < \tilde{k}_n,A_{n}\left(K_{n}\right)\cap A^{\prime}_{n}\left(K_{n}\right) ) = 0,\nonumber \\
     & \lim_{n \to \infty} P(\hat{k}_n < \tilde{k}_n) = 0.
    \label{ref39}
\end{align}

Combining \eqref{ref32} and \eqref{ref39}, we establish the desired result:
\begin{equation}
    \lim_{n \to \infty} P(\hat{k}_n = \tilde{k}_n) = 1. \nonumber
\end{equation}
\end{proof}


%
	
		\singlespacing
\bibliographystyle{econometrica-3}
\let\oldbibliography\thebibliography
\renewcommand{\thebibliography}[1]{%
  \oldbibliography{#1}%
  \setlength{\itemsep}{3pt}%
}
	{\footnotesize
		\bibliography{GSFR-BIB}
	}